\documentclass[%
    aps,
    prb,
    reprint,
    superscriptaddress
    amsfonts,
    amsmath,
    amssymb,
    eprint,
    showkeys,
    longbibliography
    twocolumn,preprintnumbers,unsortedaddress
    ]{revtex4-2}

\usepackage{graphicx}
\usepackage{dcolumn}
\usepackage{amsmath,amsbsy,amssymb,scalefnt}
\usepackage{float} 
\usepackage{lipsum} 
\usepackage{verbatim}
\usepackage{bm}
\usepackage{epstopdf}
\usepackage{amsfonts}
\usepackage{textcomp}
\usepackage[normalem]{ulem}
\usepackage{soul}
\usepackage{mathtools}
\usepackage{hyperref}
\usepackage{hyperref}
\usepackage{comment}
\usepackage{fancyhdr}
\usepackage{mathtools}
\usepackage[normalem]{ulem}
\usepackage[x11names]{xcolor}
\usepackage{orcidlink}
\hypersetup{
 pdfnewwindow=true, colorlinks=true,
 linkcolor=blue, anchorcolor=blue,
 citecolor=blue, filecolor=blue,
 menucolor=blue, urlcolor=blue
}

\sethlcolor{yellow}
\begin{document}
\title{Andreev spin qubits based on the helical edge states \\of magnetically doped two-dimensional topological insulators}

\author{Edoardo Latini\orcidlink{0009-0002-7859-9116}}
\email{edoardo.latini@polito.it}
\author{Fausto Rossi\orcidlink{0000-0001-6729-0928}}
\email{fausto.rossi@polito.it}
\author{Fabrizio Dolcini\orcidlink{0000-0001-9550-9106}}
\email{fabrizio.dolcini@polito.it}

\affiliation{Dipartimento di Scienza Applicata e Tecnologia del \href{https://ror.org/00bgk9508}{Politecnico di Torino}, I-10129 Torino, Italy}

\date{\today}

\begin{abstract}
We show that Andreev spin qubits can be realized in a Josephson junction based on the helical edge states of a two-dimensional topological insulator (quantum spin Hall system)  proximized by superconducting films, in the presence of  magnetic doping. We demonstrate that the electric  dipole transitions   between the Andreev spin states induced by the magnetic doping can be harnessed to  manipulate the Andreev spin qubit by microwave radiation pulses,  without applying an external Zeeman field or invoking ancillary states. We numerically simulate the realization of NOT and Hadamard quantum logic gates, and discuss implementations in realistic setups.
\end{abstract}

\maketitle

\section{Introduction}
Andreev spin qubits (ASQs) represent one of the most promising proposals of solid state platforms for quantum information processing. In an ASQ, a doublet of spin-split Andreev bound states (ABSs) is designed through a Josephson junction (JJ) where the weak link is characterized by spin-orbit coupling~\cite{nazarov_2003,nazarov_2010}. As compared  to qubit realizations based on conventional quantum dots, an ASQ exhibits several compelling benefits. The discrete ABSs emerge from the  superconductor gap rather than electrical confinement, making  electron-electron interaction effects  negligible. Moreover, the  readout of the ASQ quantum state can be achieved by measuring the dissipationless current flowing through the JJ~\cite{egger_2009,reynoso_2012,bosco_2025}. Finally, the spin degree of freedom is usually expected to be more robust to environmental decoherence than charge~\cite{awschalom_2001,awschalom_2007}, and spin-flip error correction architectures have also been recently proposed~\cite{fatemi_prl_2025}.

So far, implementations of ASQs have been  theoretically investigated~\cite{levy-yeyati_2017,pothier_2021,houzet-meyer_2024} and experimentally realized \cite{pothier_2019,hays_2020,hays_2021,pita-vidal_2023,kouwenhoven_2023,pita-vidal_2024,pita-vidal_2025,fatemi_prappl_2025,fatemi_prl_2025} mainly in semiconductor nanowires  with strong spin-orbit coupling contacted to two superconducting banks. Evidence of the resulting spin-split ABSs in the weak link was observed, and the possibility to control, readout and manipulate the spin qubit via microwave radiation was in principle demonstrated \cite{hays_2020,hays_2021,pita-vidal_2023,kouwenhoven_2023,pita-vidal_2024,pita-vidal_2025,fatemi_prappl_2025,fatemi_prl_2025}. 
However, in view of scalable quantum architectures a major limitation  of this platform  is the relatively short decoherence time (tens of {\rm ns})  arising from a spin noisy environment.  This is attributed to the fact that the   nanowire weak link is typically based on Indium (e.g. InAs), an element characterized by a large nuclear spin, which via hyperfine interaction acts as a spin bath causing decoherence on the ASQ~\cite{hays_2021,pita-vidal_2023,pita-vidal_2025,fatemi_prappl_2025,tahan_2025}.

During the past year, various strategies to  circumvent these difficulties have been proposed. On the one hand, it has been argued that replacing InAs nanowires with a Germanium two-dimensional hole gas  could  reduce decoherence by isotopic purification~\cite{tahan_arxiv_2025}. However, the  realization of Ge-based JJs is quite recent and, at the moment, is limited to planar junctions~\cite{katsaros_2024, defranceschi_2025}, which do not exhibit discrete ABSs. On the other hand, spin relaxation is predicted to be suppressed by shunting the ASQ 
with a capacitor, but only provided that the conditions for the Franck-Condon blockade are fulfilled~\cite{vanheck_2025}.
Other proposals for ASQs require external magnetic field~\cite{sau_2025} or ad hoc configurations such as three-terminal devices \cite{belzig_2025} or Corbino-geometries~\cite{prada-sanjose_2025}. Thus, the search for a convincing alternative to the  nanowire  implementation of ASQs   still remains a challenging open problem.

\begin{figure}[h]
\includegraphics[width=0.9\linewidth]
{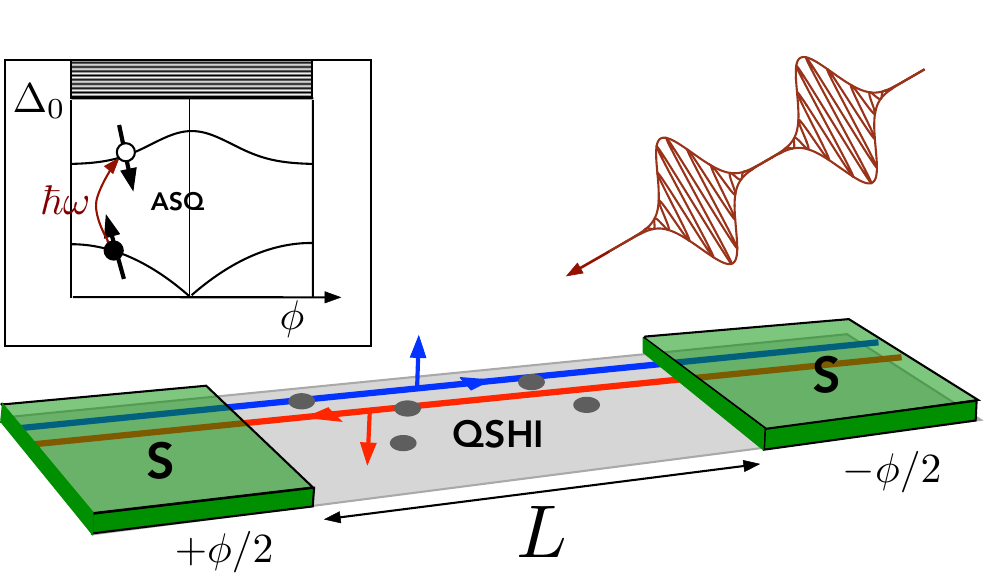}
\caption{\label{Fig1-setup} Scheme of realization  of an ASQ based on a two-dimensional topological insulator: A helical Josephson junction  realized by proximizing the   helical  edge states of a QSHI with $s$-wave superconducting (S) films.  The presence of magnetic doping (black spots) induces electric dipole transitions between the ABSs  localized in the weak link (inset), enabling the  control of the qubit via coupling with an electromagnetic radiation.}
\end{figure}
Here, we propose a quite different strategy to attack this problem,  based on  a 
 JJ   realized with a two-dimensional topological insulator, namely a quantum spin Hall insulator (QSHI)~\cite{bernevig_2006,molenkamp-zhang_review,zhang-review_2011,ando_2013}. This peculiar state of matter  has been experimentally observed in {\rm HgTe/CdTe}~\cite{zhang_2006,molenkamp_2007,brune_2010,molenkamp_2013} quantum wells, in {\rm InAs/GaSb} \cite{knez_2011,knez_2015}  bilayers,  Bi  bilayers~\cite{murakami_2006,yazdani_2014,li_book_2019},  as well as in  ${\rm WTe}_2$ monolayers~\cite{cobden_2017,shen_2017,jarillo-herrero_2018}.  Notably, not only a QSHI-based
  platform for ASQ   is realistic, it also    exhibits intrinsic advantages   as compared to current implementation relying on conventional materials. 
 First, since 
 spin-orbit coupling is  the effect underlying the topological transition in these systems, the topological edge states are helical, namely their propagation direction is locked to the spin orientation. As a consequence, the ABSs emerging in a helical JJ are naturally spin-split, providing a built-in platform to realize ASQs. Second, the  electronic  conducting states emerging at the boundaries of a QSHI are extremely robust to  perturbations. Indeed, besides being topologically protected from elastic time-reversal disorder effects~\cite{bernevig_2006,zhang_2006,molenkamp-zhang_review,zhang-review_2011,ando_2013},   they are also effectively robust to inelastic phonon coupling~\cite{budich_2012,loss-hsu_2021}, and their hyperfine interaction is  small, especially in HgTe implementations~\cite{platero_2012,rosenow_2013,loss_2017,loss_2018}. Moreover, JJs based on QSHI have already been realized in various experiments~\cite{yacoby_2013,kouwenhoven_2015,molenkamp_2016,molenkamp_natnanotech_2016,molenkamp_2024,molenkamp_2025}  and are    state-of-the-art technology by now.

The manipulation of an ASQ is customarily performed via coupling to a microwave radiation. However, as is well known, electric dipole transitions between  helical-based ABSs are a priori forbidden by selection rules imposed by their very helical nature~\cite{artemenko_2012,dora_2012, dolcini_2016}. Magnetic dipole transitions, relying on the Zeeman coupling with circularly polarized radiation~\cite{artemenko_2012,sassetti_2014}, are typically very weak and would not be suitable   to enable fast operations on an ASQ. The recently realized magnetic doping of two-dimensional topological insulators might turn the tide in this respect~\cite{yazdani_2020,molenkamp_2021,molenkamp_2024}. The effects of magnetic impurities on QSHI have been analyzed in various contexts, such as the realization of quantum anomalous Hall effect~\cite{zhang_2008,cheng_2014}, the localization of Majorana quasi-particles~\cite{dolcini_2014}, and their interplay with Coulomb interaction~\cite{zhang-bernevig_2006,maciejko_2009,tanaka_2011,johannesson_2012,altshuler_2013,cazalilla_2018}. However, as we shall show, when a QSHI is embedded in a JJ, the presence of magnetic impurities  also alters the spin texture of the ABSs, leading to nonvanishing electric dipole transition amplitudes and thereby opening up the possibility to  manipulate the ASQ with a microwave radiation.  
The scheme of the proposed setup is depicted in Fig.\ref{Fig1-setup}. A JJ, realized by proximizing a pair of counterpropagating helical  states  of a QSHI with two $s$-wave superconductors, hosts   ABS  localized within the weak link  of length~$L$ (see the inset). The ABSs are spin-nondegenerate, due to the helical nature of the edge states, and realize the ASQ. The  dark spots present in Fig.\ref{Fig1-setup} along the edge describe magnetic impurities, while    radiation pulses applied on the weak link enable  one to  control the ASQ.

The paper is organized as follows. In Sec.\ref{sec-2} we present the model for the ASQ, the resulting ABSs and their spin texture. Then, in Sec.\ref{sec-3} we analyze the coupling of the ASQ to an electromagnetic radiation, and present a detailed study of the electric dipole transition amplitude as a function of the spatial extension of the doping  and the transparency of the junction. Then, in Sec.\ref{sec-4}, we present our results about the manipulation of the ASQ, showing that it is possible to implement the NOT and Hadamard quantum gates   by  suitably applied electromagnetic pulses. We discuss our results in Sec.\ref{sec-5}, where we propose possible experimental realizations with state of the art technology,  we  address  the quantum state preparation  and the  dissipation and decoherence effects. Finally, in Sec.\ref{sec-6} we draw our conclusions.

\section{The model}
\label{sec-2}
\subsection{Hamiltonian of the system}
We consider a pair of one-dimensional counter-propagating  states flowing along the edge of a QSHI. As is customary, within the bulk gap of the topological insulator, the spectrum of the edge states can  reliably be assumed to be  linear and characterized by a Fermi velocity~$v_F$~\cite{molenkamp-zhang_review,zhang-bernevig_2006}. Moreover, the helical nature of such topological edge states implies that their spin orientation is locked to   the propagation direction. For definiteness, we shall assume that right- and left-moving electrons are characterized by spin-$\uparrow$ and spin-$\downarrow$, respectively, and are described by the  massless Dirac fermion Hamiltonian in 1+1 dimensions
\begin{equation}\label{Hhel}
\mathcal{H}_{hel} =  \int dx \, \, (\psi^\dagger_{\uparrow} , \psi^\dagger_{\downarrow} ) \, \left( v_F p_x \sigma_z -\mu\sigma_0 \right) \, \left(\begin{array}{l}  \psi^{}_{\uparrow}  \\   \psi^{}_{\downarrow} \end{array} \right),
\end{equation}
where $\psi^{}_{\uparrow}(x)$ and $\psi^{}_{\downarrow}(x)$ denote the electron field operators, $p_x=-i \hbar \partial_x$ is the momentum operator, $\sigma_z$ the Pauli matrix acting on spin space, and $\mu$   the chemical potential. 

The deposition of two $s$-wave superconducting films on the QSHI induces by proximity effect a superconducting pairing along the  two   helical regions underneath, which are separated by a mesoscopic weak link    of non-proximized helical states, with a length $L$, as depicted in Fig.\ref{Fig1-setup}. The resulting superconducting pairing  is described by the Hamiltonian term
\begin{equation}
  \mathcal{H}_{SC} =  \int dx \,\left( \Delta(x) \, \psi^\dagger_{\uparrow}   \psi^\dagger_{\downarrow} + \Delta^*(x) \psi^{}_{\downarrow} \psi^{}_{\uparrow} \right)  ,
\end{equation}
where 
\begin{equation}
\Delta(x) = \Delta_0  \left\{ \begin{array}{lcl}   \, e^{i\phi/2} & & x< -L/2, \\
0 & &| x| <  L/2, \\
\, e^{-i\phi/2} & & x>+ L/2.
\end{array}
\right.
\end{equation}
Here,  $\Delta_0>0$ is the magnitude of the induced pairing, while $\phi$ denotes the superconducting phase difference.\\

The presence of a magnetic disorder inside the weak link is accounted for by a term
\begin{equation}
\mathcal{H}_{M}=\int dx \, \, (\psi^\dagger_{\uparrow} , \psi^\dagger_{\downarrow} ) \,  \mathbf{m}(x)\cdot \boldsymbol{\sigma} \, \left(\begin{array}{l}  \psi^{}_{\uparrow}  \\   \psi^{}_{\downarrow} \end{array} \right),
\end{equation}
where $\boldsymbol{\sigma} = (\sigma_x,\sigma_y,\sigma_z)$ denotes the vector of the three Pauli matrices, and 
$\mathbf{m}(x)$ is a space-dependent magnetization vector. By construction, such term breaks time-reversal symmetry  explicitly. However, while the $m_z$ component is parallel to the natural spin orientation $z$-axis of the helical states [see Eq.(\ref{Hhel})], the orthogonal $m_x$ and $m_y$  components can induce backscattering of the helical states. As we shall see, this affects the spin texture of the  ABSs characterizing the JJ.

By   introducing the Nambu spinor $\Psi= \, (\psi^{}_{\uparrow} , \psi^{}_{\downarrow} ,  \psi^{\dagger}_{\downarrow} , -\psi^{\dagger}_{\uparrow} )^T$, the   Hamiltonian 
\begin{equation}\label{H0-def}
    \mathcal{H}_0 = \mathcal{H}_{hel} + \mathcal{H}_{SC}+ \mathcal{H}_{M}
\end{equation}    
can be rewritten, up to a constant, in the Bogoliubov de Gennes (BdG) form 
$
\mathcal{H}_0 =(1/2) \int \; \Psi^\dagger(x) H_{\rm BdG}(x) \Psi^{}(x)\, dx $, 
where 
\begin{equation}
H_{\rm BdG}(x)= \begin{pmatrix}
 H_{e}(x)   &   \Delta(x) \sigma_0       \\  
 & \\
 \Delta^*(x) \sigma_0   &  H_{h}(x) \end{pmatrix} \quad \label{H-BdG} 
\end{equation}
is the BdG Hamiltonian, with 
\begin{subequations}
\begin{align}
H_{e}(x) &= v_F \sigma_z p_x -\mu \sigma_0 + \mathbf{m}(x)\cdot \boldsymbol{\sigma}, \\
H_{h}(x) &=-\mathcal{T} H_{e} \mathcal{T}^{-1} = -\sigma_y (H_{e})^* \sigma_y= \nonumber \\
& =-v_F \sigma_z p_x +\mu \sigma_0 + \mathbf{m}(x)\cdot \boldsymbol{\sigma}\; 
\end{align}
\end{subequations}
denoting the electron(e) and hole(h) sector diagonal blocks, respectively. Here,  $\sigma_0$ is  the $2 \times 2$ identity matrix in spin space, and $\mathcal{T} = - i\sigma_y K $ is the time-reversal operator, with $K$ denoting the complex conjugation.  
By re-expressing the Nambu field $\Psi$
in terms of the eigenfunction set $\Phi_i(x)$  of the BdG Hamiltonian (\ref{H-BdG})  
\begin{equation}\label{Psi-gamma_n-compact}
 \displaystyle     {\Psi}(x) = \sum_{E_i \ge 0} \left( \Phi^{(+)}_i(x)  {\gamma}^{}_i + \Phi^{(-)}_i(x)  {\gamma}^{\dagger}_i\right) ,
\end{equation}  
the Hamiltonian $\mathcal{H}_0$ can be brought to the diagonal form of a  two-level system collection 
\begin{equation}\label{H-inhomo-gamma}
 \displaystyle  \mathcal{H} = \sum_{E_i \ge 0} E_i(\phi) \left(  {\gamma}^\dagger_i {\gamma}^{}_i \\  -\frac{1}{2}\right) ,
\end{equation} 
with $\gamma_i$ denoting   fermionic Bogoliubov quasi-particle operators. In Eq.(\ref{Psi-gamma_n-compact}), $\Phi_i^{(+)}(x)   =\left( 
 {u}_{\uparrow,i}  ,  {u}_{\downarrow,i},
 {v}_{\downarrow,i}  , {v}_{\uparrow,i}   \right)^T(x)  
$ 
denote the eigenfunctions of the BdG Hamiltonian (\ref{H-BdG}) with positive eigenvalues $E_i \ge 0$ (discrete and continuous spectrum), while $\Phi_i^{(-)}(x) = (\tau_y \otimes \sigma_y) (\Phi_i^{(+)}(x))^*$ are the charge-conjugated of $\Phi_i^{(+)}$, and  have energies $-E_i$.
\subsection{Andreev bound states}
Due to the helical nature of the QSHI edge states, the  ABSs, labeled by $i=1,\ldots N$, are non degenerate in spin. They are characterized by a discrete subgap spectrum $0 \le E_i(\phi) < \Delta_0$ and by wavefunctions $\Phi_i(x)$ localized in the weak link. As is well known, the ABSs can be determined by combining the particle-hole Andreev scattering at the interfaces $x=\pm  L/2$ with the normal scattering induced by the magnetic  doping present in the weak link~\cite{beenakker_1991}. Indeed any   static magnetic disorder  with {\it arbitrary}  profile $\mathbf{m}(x)$ can be described in terms of a unitary scattering matrix. Specifically, the  electron scattering matrix $S_{e}(E)$, characterized by transmission  amplitudes $t_e, t^\prime_e$ and  reflection amplitudes $r_e, r^\prime_e$,  can always be expressed in the form
{\small
\begin{eqnarray}
\lefteqn{S_{e}(E) =\,\left( 
\begin{array}{lcl} 
r_e(E)  &  t^\prime_e(E)  \\ & \\
t_e(E)  & r^\prime_e(E) 
\end{array} \right) =} & &  \label{S0-e-gen} \\ &=& e^{i\Gamma_m(E)} \!   
\begin{pmatrix} 
-i  \,e^{+i\Theta_m(E)}   \,   \sqrt{1-T_E}  &    \,e^{i \chi_m(E)} \,  \sqrt{T_E}  \\ & \\
   \, e^{-i \chi_m(E)} \,  \sqrt{T_E} &-i  \,e^{-i\Theta_m(E)}   \,   \sqrt{1-T_E} 
\end{pmatrix} \,, \nonumber 
\end{eqnarray}
}

\noindent where  $T_E=|t_e|^2=|t^\prime_e|^2$ and $R_E=1-T_E$ denote the energy-dependent transmission and reflection coefficients, respectively, while $\Gamma_m$, $\Theta_m$ and $\chi_m$ are complex phase parameters.
For holes, the scattering matrix   is obtained as
\begin{equation}\label{Sh-Se}
    S_{h}(E)=-\sigma_z \, S_{e}^*(-E)\, \sigma_z .
\end{equation}
The equation for the ABS energy levels  is~\cite{dolcini_2014}
\begin{equation}\label{ABS-gen-fin}
\begin{aligned}
    \cos^2 &\left[ \arccos\frac{E}{\Delta_0}-\frac{E L}{\hbar v_F} -\Gamma_m^A(E)\right]\\ 
    &=\frac{1}{2} \left( 1- \sqrt{R_{E} R_{-E}} \, \cos\left(   2 \Theta_m^A(E)\right)\right.\\
    &\left.\quad+ \sqrt{T_{E} T_{-E}}  \, \cos(\phi-2\chi^S_m(E))  \right)  
\end{aligned}
\end{equation}
where
\begin{equation}
\begin{aligned}
    \Theta_m^A(E) &= \frac{1}{2} (\Theta_m(E)-\Theta_m(-E)),   \\
    \Gamma_m^A(E) &= \frac{1}{2} (\Gamma_m(E)-\Gamma_m(-E)),  \\
    \chi_m^S(E) &= \frac{1}{2} (\chi_m(E)+\chi_m(-E))    
\end{aligned} 
\end{equation}\label{PhiA-GammaA-chiS-bis}
are determined by the
parameters characterizing the    scattering matrix in Eq.(\ref{S0-e-gen}).
\\

As compared to the   studies on JJs  aiming to determine the current-phase relation of the Josephson current,  for the present problem of   ASQ realizations  two important aspects are worth noticing.
Firstly, designing a  qubit requires to deal with a limited number of ABSs only. Indeed, while at  least two ABS are needed to realize a two-level system, the presence of too many levels would be detrimental, for their energy separation would be too small to enable the selective  manipulation of the quantum state with a radiation. The number of ABS is known~\cite{bagwell_1992} to depend on the ratio
\begin{equation}\label{lambda-def}
    \lambda=\frac{L}{\xi_S} = \frac{\Delta_0}{\hbar v_F/L}
\end{equation}
of the  weak link length $L$   to the superconducting coherence length $\xi_S=\hbar v_F/\Delta_0$. Typically, analytical results can be obtained in the limits of either short junction ($L \ll \xi_S$)~\cite{beenakker_1991_bis}, where only one ABS is present, or long junction ($L \gg \xi_S$)~\cite{ishii_1970}, where many closely spaced energy levels are present. For an ASQ, however, one has to operate in the {\it intermediate} length regime, where $L \sim \xi_S$, and determine the ABS   spectrum $E_i(\phi)$ by  solving Eq.(\ref{ABS-gen-fin}) numerically.  
The second relevant aspect is that, while the Josephson current only requires to find the eigenvalues $E_i(\phi)$, modelling an ASQ  also implies to determine the ABS {\it eigenfunctions} $\Phi_i(x)$. 
Indeed the  manipulation of the ASQ crucially depends on the possibility to induce spin-flip processes, i.e. to control transitions between the two ASQ states. As we shall discuss below in details, in the QSHI helical states,  where  charge current is closely related to spin, such transitions are determined by   the overlap integrals of the current between pairs of different ABS eigenfunctions $\Phi_i(x)$'s, which in turn are closely related to the matrix entries of the electric dipole. Thus, the manipulation of the ASQ via the coupling to an electromagnetic radiation relies on the possibility to have nonvanishing electric dipole transitions between the ABS. Notably, selection rules play a crucial role in this mechanism.

Indeed, for a clean JJ, the helical nature of the QSHI edge states implies that each ABS  wavefunction $\Phi_i(x)$ is an eigenfunction of the $z$-component of the spin operator. One has either spin-$\uparrow$ ABS, i.e.  excitations characterized by the combination of spin-$\uparrow$ particles and spin-$\downarrow$ holes, or spin-$\downarrow$ ABS, i.e.  excitations characterized by the combination of spin-$\downarrow$ particles and spin-$\uparrow$ holes. In this case,  spin  selection rules forbid  any electric dipole transitions between opposite spin states, and it is not possible to manipulate the ASQ. 

\begin{figure}
\includegraphics[width=0.9\linewidth]
{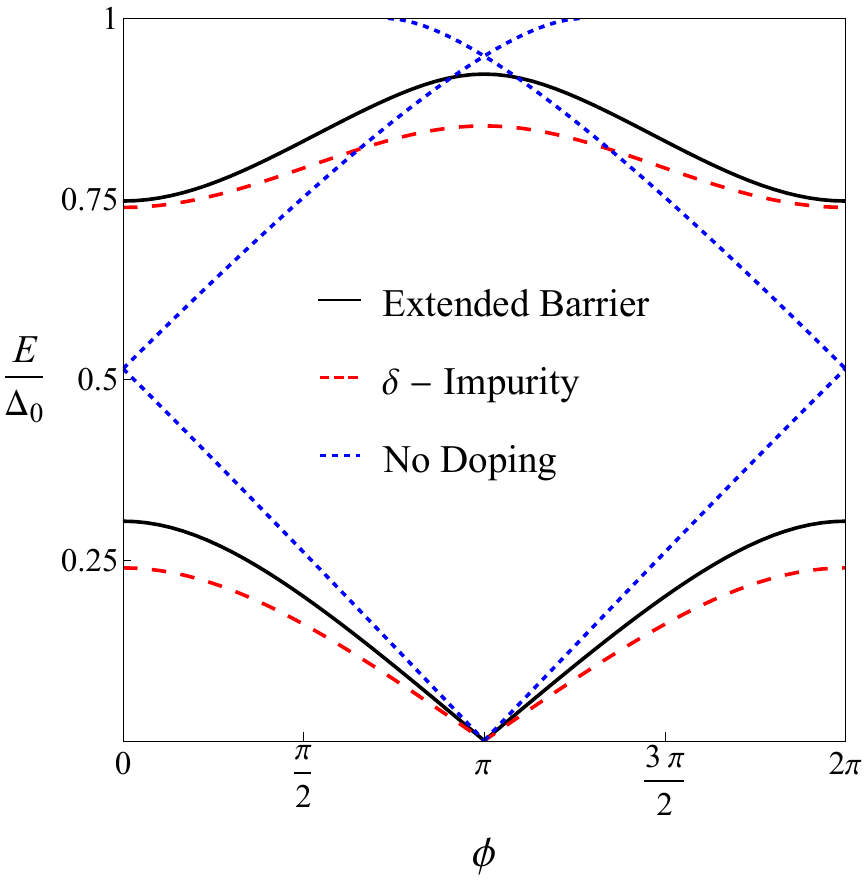}
\caption{\label{Fig2-ABS}   The ABS energy levels as a function of the superconducting phase difference $\phi$ are shown for a helical JJ with a length parameter $\lambda=2$, in the presence of   magnetic barrier, characterized by a parameter $\alpha=m_\perp L_m/\hbar v_F=1$, for chemical potential  value $\mu= \Delta_0/2$. The black solid curve and the dashed red curve describe the cases of   magnetic disorder extended over the entire weak link ($L_m \rightarrow L$) and of a localized magnetic impurity ($L_m \rightarrow 0$) located at $x_0=L/6$, respectively. For comparison, the dotted blue curve depicts the ABS in the absence of any magnetic disorder. }
\end{figure}

The presence of magnetic disorder   is the mechanism to circumvent this problem. In order to illustrate how, we first observe that, in terms of spatial distribution, one can  have  various disorder scenarios, ranging from an extremely dilute disorder, where at most one magnetic impurity occurs within the weak link of the JJ, to a spatially dense  disorder, where the weak link hosts  a uniform  distribution of impurities with  randomly oriented magnetizations. Here, our purpose is not to carry out a detailed investigation on the possible disorder configurations or to compare a single sample vs a sample-averaging. Rather, we aim to illustrate the basic ingredients  leading to manipulate the ASQ in presence of magnetic disorder.
Specifically, the   $m_x$ and $m_y$  components   that are {\it orthogonal} to the natural spin orientation axis   alter the spin texture of the ABS  enabling  spin-flip  transitions.  The simplest model to illustrate this effect is the ``magnetic barrier'' profile 
\begin{equation}\label{barrier}
\mathbf{m}(x)\cdot \boldsymbol{\sigma}= \left\{ \begin{array}{lcl} m_\perp \sigma_x,    & & x_0-\frac{L_m}{2} < x <x_0+\frac{L_m}{2}, \\ & & \\
0, & & \mbox{\small otherwise,}\end{array} \right. 
\end{equation}
where $L_m$ and $x_0$ denote the length and the center of the magnetic barrier inside the weak link, respectively, while  the  additional  conditions $-L/2<x_0-L_m/2$ and $x_0+L_m/2<L/2$  ensure that the barrier is inside the weak link. Two   cases are noteworthy. The limit of a narrow and high barrier, $L_m \rightarrow 0,  m_\perp \rightarrow \infty$, with keeping the ``area" parameter $\alpha=m_\perp L_m/\hbar v_F$ constant, describes the case of  spatially dilute disorder, namely  one single $\delta$-like   magnetic  impurity located at a generic position~$x_0$. In contrast,   the case $L_m \rightarrow L$ (with $x_0 \rightarrow 0$) can effectively mimic  the case of  magnetic disorder uniformly distributed along the entire weak link   (see at the end of this section for more details). For these two limiting cases, the two ABS present in an intermediate length Josephson junction ($L=2 \xi_S$)  are shown by the solid black curve and the dashed red curve depicted in Fig.\ref{Fig2-ABS}. For a comparison, the dotted blue curve illustrates the ABS in the absence of any magnetic disorder (clean helical JJ). As one can see, the zero-energy crossing of the lower ABS at $\phi=\pi$ is robust  to the presence of magnetic disorder. This is known to be a hallmark of a topological JJ~\cite{yakovenko_2004,fu-kane_2009,beenakker_2013,sassetti_2013,trauzettel_2014}  and is in striking contrast with the behavior of conventional JJs~\cite{bagwell_1992}.

We emphasize that the ABS wavefunctions $\Phi_1$ and $\Phi_2$ are mutually {\it orthogonal}, since they are eigenfunctions of the BdG Hamiltonian (\ref{H-BdG}) related to two different eigenvalues $E_1$ and $E_2$. Thus, they represent the quantum computational states $|0\rangle $ and $|1\rangle$ of the ASQ. 
At a given superconducting phase bias $\phi$, the energy difference between the two ABSs    determines the frequency $\omega=(E_2-E_1)/\hbar$ of the radiation enabling the  manipulation of the ASQ state. As one can see from Fig.\ref{Fig2-ABS}, such energy difference   is maximal for $\phi=\pi$ and minimal at phase difference $\phi=0, 2\pi$. This might at first suggest that the optimal value of $\phi$ for  ASQ manipulation is $\pi$. However, as we shall see in the next section,  the   electric dipole  transition amplitudes  depend on $\phi$ in a way   opposite to the energy gap. Both these aspects have to be taken into account in investigating the ASQ   manipulation.\\

\begin{figure}
\includegraphics[width=\linewidth]
{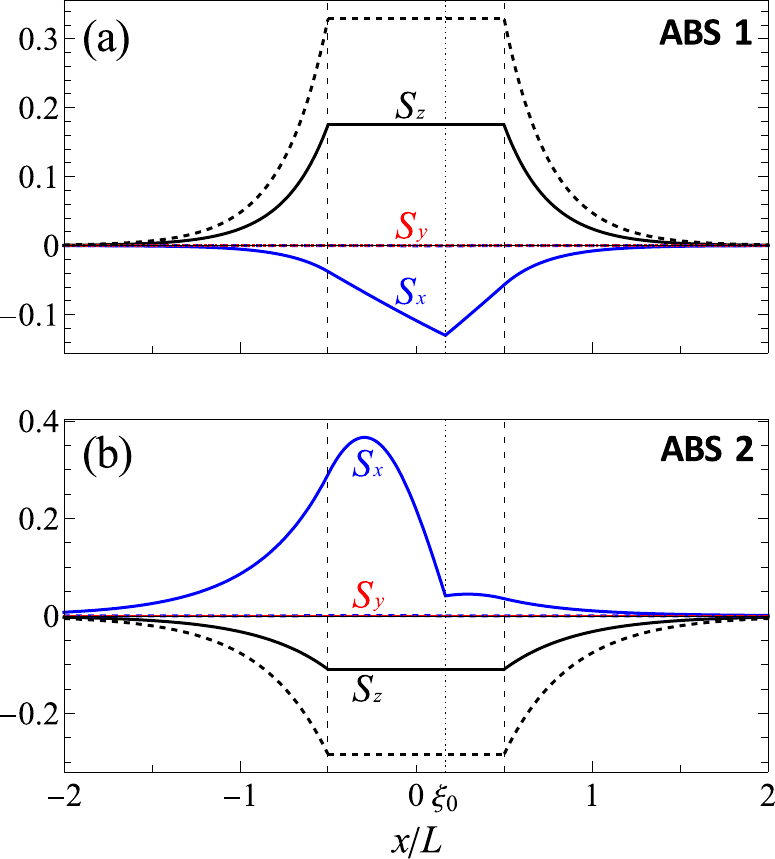}
\caption{\label{Figure-03-spin-density}  Spatial profile of the three spin density components $S_x, S_y, S_z$ (in units of $1/L$), for the two ABSs present in a weak link with length parameter $\lambda=2$.  The vertical dashed lines at $x/L=\pm 1/2$ are a guide to the eye to identify the  weak link region. The value of the superconducting phase difference is $\phi=\pi/2$. The solid curves refer to the case of the presence of a  magnetic $\delta$-impurity located at $\xi_0=x_0/L=1/6$ and with   parameter $\alpha=1$, while the  dashed curves depict, for comparison, the case of a clean JJ without magnetic disorder, where $S_z$ is the only nonvanishing component.  
}
\end{figure}

For each discrete ABS eigenvalue $E_i(\phi)$, we have determined the related   eigenfunction $\Phi_i(x)$.  While details about the  explicit computation are given in the Appendix~\ref{AppA}, here we would like to discuss  the spatial profile of their spin density, defined as
\begin{equation}
 S^{(i)}_\alpha(x)=   \Phi^\dagger_i(x)  \Sigma_\alpha \Phi_i(x),    \hspace{1cm} \alpha=x,y,z, 
\end{equation}
where $\Sigma_\alpha=\tau_0 \otimes \sigma_\alpha$ are the spin-operators  in Nambu formalism. The result for the case of a JJ with two ABSs $\Phi_1$ and~$\Phi_2$ is  illustrated in the two panels (a) and~(b) of Fig.\ref{Figure-03-spin-density}, respectively. In each panel, the dashed vertical dotted lines at $x/L =\pm 1/2$ are a guide to the eye to identify the JJ weak link, and the three solid curves describe the three components $S_\alpha$ of the spin density, in the presence of a magnetic $\delta$-impurity located at $\xi_0=x_0/L$. For a comparison, the   dashed curves describe the case of a clean JJ, where $S_z$ is the only nonvanishing component, since the ABS are eigenstates of $\Sigma_z$.  As one can see, the presence of the magnetic impurity alters the spin texture of the ABS, decreasing the magnitude of the $S_z$ component and introducing a $S_x \neq 0$ component, which exhibits a singularity at the impurity location $\xi_0$  (marked by a vertical thin dotted line).  A scheme to detect the the helical state spin texture has been  proposed in Ref.\cite{trauzettel-calzona_2022}.

Before concluding this section, we would like to illustrate why  the   profile (\ref{barrier}) is expected to effectively describe  the case of   uniformly distributed disorder, when $L_m \rightarrow L$.
We start by observing that, 
  in the limit of one single isolated impurity, the specific orientation of the magnetization $\mathbf{m}$ within the $x$-$y$ plane  orthogonal to the natural spin orientation of the helical states  is irrelevant: It is just equivalent to a rotation  of the spin axes around $z$.  In the case one has   a few
impurities,  quantum interference phenomena between the backscattering processes at the various impurites might arise, depending on the relative impurity distance and the relative orientation of their magnetization in the $x$-$y$ plane~\cite{dolcini_2014}. However,
if one considers the case of many  impurities that are uniformly distributed along the weak link with random magnetization directions, the quantum interference effects would mutually cancel out, and only the classical (i.e. incoherent) contributions -consisting in the sum of the effects of each individual impurity- would matter. Because for each individual impurity the magnetization  orientation is irrelevant, the effective result is expected to be the same when all impurity magnetizations are aligned in the same direction. Thus, a uniform distribution of weak impurities can be mimicked by the uniform magnetization profile (\ref{barrier}).

\section{Coupling to electromagnetic radiation and  transition amplitudes}
\label{sec-3}
We now consider a full Hamiltonian
\begin{equation}\label{Hfull}
\mathcal{H}=\mathcal{H}_0+\mathcal{H}^\prime, 
\end{equation}
where $\mathcal{H}_0$ is given by Eq.(\ref{H0-def}), while the additional   time-dependent term 
\begin{equation}\label{Hprime-def-pre}
\mathcal{H}^\prime=-\int dx \mathcal{J}(x)\, A(x,t)
\end{equation}
describes the coupling of the helical states to an electromagnetic radiation. Here, $A(x,t)$ is the vector potential describing an electric field pulse $E(x,t)=-\partial_t A(x,t)$ in Système International (SI) units, while
\begin{eqnarray}
\mathcal{J}(x) &=& {\rm e}v_F \left(\psi^\dagger_\uparrow(x)\psi_\uparrow(x)- \psi^\dagger_{\downarrow}(x)     \psi_{\downarrow}(x)\right) =\nonumber \\
&=& \frac{{\rm e}\, {v_F}}{2} \Psi^\dagger(x) (\tau_0 \otimes \sigma_z)  \Psi^{}(x) \label{J-def}
\end{eqnarray}
is the current density operator, with ${\rm e}$ denoting the elementary charge and  $\tau_0$  the $2 \times 2$ identity in Nambu space. Equation~(\ref{J-def}) explicitly shows that the current operator  contains the  $z$-component  $\Sigma_z=\tau_0 \otimes \sigma_z$ of the spin-operator  in Nambu formalism. This implies that, differently from conventional materials,  the spin texture of the ABS realized with QSHI edge states matters in determining  inter-level  transitions  induced by the coupling (\ref{Hprime-def-pre}).\\

We shall now make the following assumptions. First, since the sub-${\rm THz}$ frequency range  related to the typical superconducting gap  $\Delta_0 \lesssim 1\,{\rm meV}$ corresponds to a  wavelength of a few ${\rm mm}$, i.e. far larger than the typical length $L \lesssim 1\,\mu{\rm m}$ of the junction, the electromagnetic pulse can be assumed to be spatially  uniform, and the coupling term (\ref{Hprime-def-pre}) can be approximated to
\begin{equation}\label{Hprime-def}
\mathcal{H}^\prime \simeq -A(t)\int dx \, \mathcal{J}(x)\, .
\end{equation}
The second assumption is that, since  the  physical processes relevant to the ASQ   occur  in the energy range $0 \le E<\Delta_0$  within the superconducting gap $\Delta_0$, the expansion (\ref{Psi-gamma_n-compact}) of  the electron field operator in terms of all the eigenfunctions of the Bogoliubov de Gennes Equations can be well approximated by retaining only the $N$ discrete ABS solutions, and neglecting the contribution from the continuum states ($E>\Delta_0$)
\begin{equation}\label{Psi-gamma_n-compact-only-ABS}
 \displaystyle     {\Psi}(x) \simeq \sum_{0 \le E_j \le \Delta_0} \left( \Phi^{(+)}_j(x)  {\gamma}^{}_j+ \Phi^{(-)}_j(x)  {\gamma}^{\dagger}_j\right)  .
\end{equation}
This reduces the dimensionality of the problem to the essential degrees of freedom needed for the ASQ    manipulation. 
Inserting Eq.(\ref{Psi-gamma_n-compact-only-ABS}) into Eq.(\ref{J-def}), one can re-express the coupling $\mathcal{H}^\prime$  in the Bogoliubov quasi-particle basis that diagonalizes  the $\mathcal{H}_0$ term    [see Eq.(\ref{H-inhomo-gamma})], obtaining 
\begin{eqnarray}\label{Hprime-in-terms-of-gamma-v2}
\mathcal{H}^\prime  &=&  - \frac{{\rm e} v_F}{2}   A(t)\!\!\! \!\! \sum_{0 \le \, E_i,  E_j < \Delta_0}\!\! \left( g_{ij}  \gamma^\dagger_i \gamma^{}_j-g^*_{ij}  \gamma^{}_i \gamma^{\dagger}_j+ \right.   \nonumber \\
& & \left. \hspace{3cm} +\tilde{g}_{ij}  \gamma^{\dagger}_i \gamma^{\dagger}_j+\tilde{g}^*_{ji}  \gamma^{}_i \gamma^{}_j\right)  ,
\end{eqnarray}
where 
\begin{eqnarray}
g_{ij}   = \int dx \left(  \Phi^{(+)}_i(x)    \right)^\dagger(\tau_0 \otimes \sigma_z) \Phi^{(+)}_j(x)\,\,  \label{g_ij-def} 
\end{eqnarray}
and
\begin{eqnarray}
\tilde{g}_{ij}  =  \int dx \left(  \Phi^{(+)}_i(x)    \right)^\dagger((-i\tau_y) \otimes \sigma_x) \left(\Phi^{(+)}_j(x)\right)^*\,\,  \, \label{gtilde_ij-def}
\end{eqnarray}
are dimensionless overlap integrals. The $g_{ij}$ coefficients in Eq.(\ref{g_ij-def}), which   by definition are bounded ($|g_{ij}| \le 1$) and   fulfill the property $g_{ji} = g^*_{ij}$, identify  the matrix entries of the current density operator (\ref{J-def}), divided by ${\rm e} v_F$, between  ABSs. 
In particular, a diagonal entry $g_{ii}$ is related to the spatially averaged current carried by the $i$-th ABS, while the off-diagonal entries $g_{ij}$ (with $i \neq j$) describe  the processes of photoexcitation from the $j$-th ABS to the $i$-th ABS. 
Notably, despite the expression~(\ref{J-def}) of the QSHI edge state current density is independent of the momentum $p_x$ and is therefore quite different from 
the case of conventional materials with a parabolic spectrum such as nanowires, one can   show   that the off-diagonal entries $g_{ij}$ (with $i \neq j$) are related to 
  the electric dipole operator,  which in Nambu formalism acquires the form $d={\rm e} \, x\, (\tau_z \otimes \sigma_0)/2$. To this purpose, it is sufficient to realize that the major contribution to the $g_{ij}$ coefficients stems from the weak link region where the ABS are localized. Inside the weak link one has   $\left[H_{\rm BdG}(x), d\right]=-i {\rm e} \hbar v_F (\tau_0 \otimes \sigma_z)/2= -i \hbar J$. Applying the ABS wavefunctions $\Phi^\dagger_i$ (on the left) and $\Phi^{}_j$ (on the right) of this relation and exploiting $H_{\rm BdG}\Phi_i=E_i \Phi_i$, one finds 
\begin{eqnarray}
 \Phi_i^\dagger   (\tau_0 \otimes \sigma_z) \,\Phi^{}_j =  \frac{2 i (E_i-E_j)}{{\rm e}\hbar  v_F}\Phi_i^\dagger   \,{d}\,  \,\Phi^{}_j.
\end{eqnarray}
 For this reason, the ASQ spin-flip processes determined by $g_{ij}$ are electric dipole induced spin processes. 
 
As far as the   $\tilde{g}_{ij}$ are concerned,  they are also bounded ($|\tilde{g}_{ij}| \le 1$),  fulfill $\tilde{g}_{ji} = -\tilde{g}_{ij}$, and represent  the matrix entries of Eq.(\ref{J-def}) between the charge conjugated   of the $j$-th ABS and the $i$-th  ABS. The $\tilde{g}_{ij}$ physically describe photoexcitation from the superconducting condensate to a pair $(i,j)$ of ABSs, as can be deduced from inspection of Eq.(\ref{Hprime-in-terms-of-gamma-v2}).

\subsection{Single particle density matrix in Nambu formalism}
The quantum state of the ASQ can be characterized by the $2N \times 2N$ single particle density matrix in Nambu formalism  
\begin{equation}\label{rho_sp-def}
\boldsymbol{\rho}=\left\langle \boldsymbol{\Gamma}^\dagger \otimes \boldsymbol{\Gamma}\right\rangle = \begin{pmatrix} 
\rho^{ee} & \rho^{eh}\\
\rho^{he} & \rho^{hh}
\end{pmatrix},
\end{equation}
where the entries of the  diagonal particle and hole blocks  are
\begin{equation}
    \rho^{ee}_{ij}=\langle \gamma^\dagger_j \gamma^{}_i \rangle, \hspace{2cm}\rho^{hh}_{ij}=\langle \gamma^{}_j \gamma^{\dagger}_i \rangle , 
\end{equation}
while for the anomalous off-diagonal blocks one has
\begin{equation}\label{rho-eh-def}
\rho^{eh}_{ij}=    
\langle  \gamma^{}_j\gamma^{}_i  \rangle, \hspace{2cm} 
\rho^{he}_{ij}=    
\langle  \gamma^{\dagger}_j\gamma^{\dagger}_i  \rangle  ,
\end{equation}
 with $i,j=1, \ldots N$ (number of ABSs).
The  fermionic nature of the operators $\gamma_i$'s implies that   $\rho^{eh}_{ii}=\rho^{he}_{ii}=0$.\\

It is possible to show that the single particle density matrix defined in Eq.(\ref{rho_sp-def}) fulfills a Liouville -- von Neumann Equation governed by the   BdG Hamiltonian in the $\gamma$-Bogoliubov quasi-particle basis. 
Explicitly, in the interaction(${\rm I}$) picture, one has
\begin{equation}\label{LvN-Eq-Interaction}
i \hbar \frac{d\boldsymbol{\rho}_{\rm I}}{dt} = \left[ H^\prime_{\rm I}(t)\, ,\,\boldsymbol{\rho}_{\rm I}(t) \right], 
\end{equation}
where
\begin{equation}\label{Hprime-Int-pic}
{H}^\prime_{\rm I}(t)= - {\rm e} v_F A(t) \begin{pmatrix}
     \mathsf{g}^{ee}_{\rm I}(t) & \tilde{\mathsf{g}}^{eh}_{\rm I}(t) \\ & \\
\tilde{\mathsf{g}}^{he}_{\rm I}(t) & \mathsf{g}^{hh}_{\rm I}(t)
 \end{pmatrix}
\end{equation}
is the perturbing  Hamiltonian, with
\begin{eqnarray}
\left(\mathsf{g}^{ee}_{\rm I}(t)\right)_{ij} &=& g_{ij}\, e^{i\frac{(E_i-E_j)t}{\hbar}},  
 \\(\mathsf{g}^{hh}_{\rm I}(t))_{ij}&=&-g^*_{ij}\, e^{-i\frac{(E_i-E_j)t}{\hbar}},\\
 \left(\tilde{\mathsf{g}}^{eh}_{\rm I}(t)\right)_{ij} &=&  \tilde{g}_{ij}\, e^{i\frac{(E_i+E_j)t}{\hbar}},   \\(\tilde{\mathsf{g}}^{he}_{\rm I}(t))_{ij}&=&\tilde{g}^*_{ji}\, e^{-i\frac{(E_i+E_j)t}{\hbar}} .
\end{eqnarray}
Here,  
$g_{ij}$ and $\tilde{g}_{ij}$ are given by Eqs.(\ref{g_ij-def}) and (\ref{gtilde_ij-def}), respectively, while the time-dependent phase factors  are related to the ABS energy levels $E_i$'s characterizing the $2N \times 2N$  unperturbed Hamiltonian
\begin{equation}\label{H0-and-Hprime-blocks-v2}
 {H}_0=  \begin{pmatrix}
      {H}^{ee}_{0} & 0 \\
0 &  {H}^{hh}_{0} 
 \end{pmatrix} ,
\end{equation}
which is by definition block-diagonal with particle and hole blocks given by 
\begin{equation}\label{H0-Bogolubov}
 {H}^{ee}_{0}= \begin{pmatrix}
     E_1 & 0 &   \ldots \\
     0 & E_2 &   \ldots  \\
   0 & 0 &   \ddots   
 \end{pmatrix},   \hspace{1cm} {H}^{hh}_{0}=- {H}^{ee}_{0}.
\end{equation}
  
\subsection{Electric dipole  transition amplitudes}
The structure of the radiation coupling term (\ref{Hprime-Int-pic})  shows that the   diagonal  coefficients $g_{ii}$ lead to a renormalization of the unperturbed ABS eigenvalues in Eq.(\ref{H0-Bogolubov}), which a priori could affect the resonance frequency for the  inter-ABS  transitions.  
Moreover,   the $\tilde{g}_{ij}$ coefficients couple the electron-hole sectors in  Eq.(\ref{Hprime-Int-pic}). 
Although our simulations    to be presented in the next section fully take  into  account   the ${g}_{ii}$ and $\tilde{g}_{ij}$ coefficients too,  we anticipate
that both these effects turn out to be negligible, at least in the prototypical case of a JJ hosting two ABSs, as we shall discuss at the end of Sec.\ref{sec-4}. 
The key ingredient to induce the    spin-flip  transitions and to enable the ASQ   manipulation   is the {\it off-diagonal} coefficients $g_{ij}$ (with $i \neq j$),  i.e. the electric dipole transition entries, which   we would like to discuss here. \\

\begin{figure}
\includegraphics[width=\linewidth]
{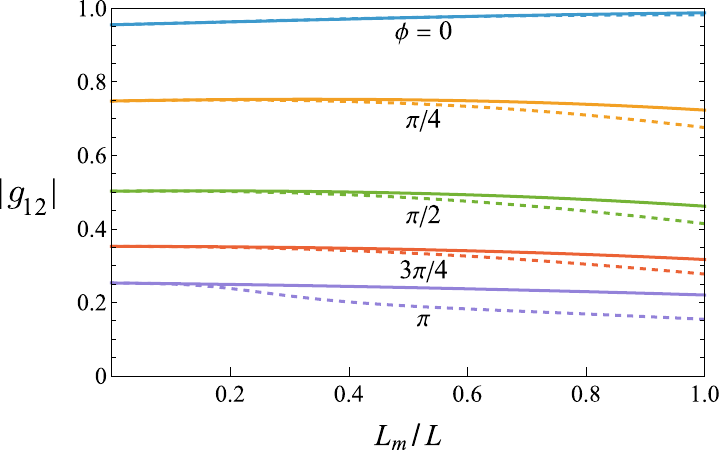}
\caption{\label{Fig4-g12}   The magnitude $|g_{12}|$ of the  electric-dipole transition amplitude between two ABSs is shown as a function of the  ratio of the spatial extension $L_m$ of the magnetic disorder to the JJ weak link length $L$, for a fixed value of the magnetic barrier parameter $\alpha=m_\perp L_m/\hbar v_F =0.5$, and for different values of the superconducting phase difference $\phi$. Solid curves refer to chemical potential   $\mu=0$, while dashed curves to $\mu=\Delta_0/2$. The magnetic disorder is distributed around the center $x_0=0$ of the weak link, which has a   length parameter  $\lambda=2$.  }
\end{figure}

For definiteness, we focus  on the  case of a JJ characterized by two ABSs, obtained by setting the weak link length parameter to $\lambda=2$ [see Eq.(\ref{lambda-def})]. By considering the presence of a magnetic disorder barrier Eq.(\ref{barrier}) with energy height $m_\perp$ and length $L_m$  inside the weak link ($0\le L_m \le L$),   we have investigated how the    electric dipole  transition amplitude $g_{12}$ between the two ABSs depends on the various parameters of the barrier and the junction.   
Figure~\ref{Fig4-g12} analyzes the dependence of $g_{12}$ on the spatial extension $L_m$ of the barrier,  for a fixed value of the barrier ``area" parameter $\alpha=L_m m_\perp/\hbar v_F$. Solid curves illustrate the behavior for a vanishing chemical potential $\mu=0$, and show that $g_{12}$ is roughly independent of the  spatial extension $L_m$ of the barrier.
  The various curves   refer  to different values of the superconducting phase difference~$\phi$. In particular, one has $|g_{12}| \lesssim 1$ for vanishing phase difference ($\phi=0)$, while  $|g_{12}|$ decreases with increasing $\phi$, and reaches its minimal value for $\phi = \pi$. 
The dashed curves  describe the case  $\mu=\Delta_0/2$. As one can see, a finite value   $|\mu| < \Delta_0$   of the chemical potential leads  to a slight reduction of  $g_{12}$  only for $L_m \sim L$,    whereas  $g_{12}$  is independent of the value of   $\mu$ for $L_m \ll L$. Notice also that such a reduction depends on $\phi$ as well, with the curves at $\phi=0$ being insensitive to the variations of $\mu$.   In fact, it can be shown that only far away from the Dirac point $|\mu| \gg \Delta_0$ and for phase differences $\pi/2 < \phi \lesssim \pi$, the dependence of~$g_{12}$ on $L_m/L$ becomes significant. 
 These results suggest  that, operating sufficiently close to the Dirac point ($|\mu| < \Delta_0$), a single $\delta$-like impurity and a uniformly distributed barrier with equal barrier ``area" $\alpha \lesssim 1$ yield  a quite similar  electric dipole  transition amplitude.

\begin{figure}
\includegraphics[width=\linewidth]
{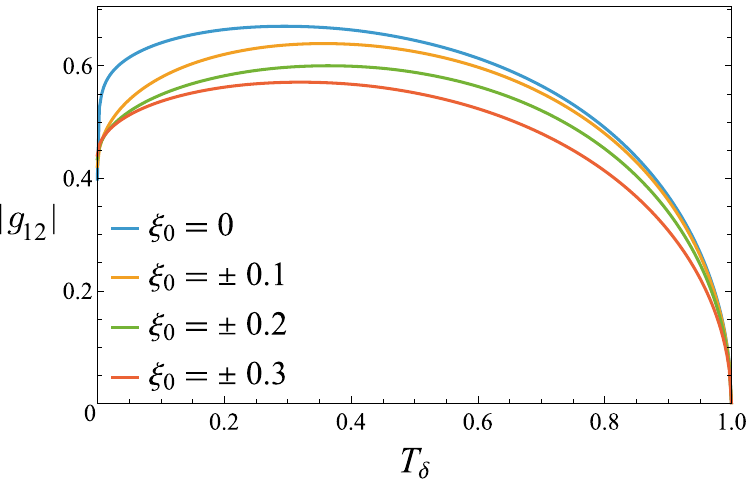}
\caption{\label{Fig5-g12-afo-Tdelta}    The magnitude $|g_{12}|$ of the electric dipole transition amplitude between two ABSs is shown as a function of the transmission $T_\delta$ of a magnetic $\delta$-impurity present in the weak link, for different values of the impurity relative location  $\xi_0=x_0/L$ from the weak link center. The  length parameter is $\lambda=2$, and the superconducting phase difference is $\phi=\pi/2$.}
\end{figure}

For this reason, we now  focus  on the case of a  magnetic $\delta$-impurity, whose transmission coefficient  acquires a simple expression $T_E=T_\delta=1/\cosh^2{\alpha}$ that is independent of the energy and of the chemical potential $\mu$. 
Figure~\ref{Fig5-g12-afo-Tdelta} displays $|g_{12}|$   as a function of the transmission coefficient $T_\delta$, for various values of the impurity position within the weak link, for a superconducting phase difference $\phi=\pi/2$.  
Two features are noteworthy. First,
 in the limit $T_\delta \rightarrow 1$, one can see that  $|g_{12}|\rightarrow 0$ for all curves. Indeed in this limit, which corresponds to the absence of disorder, the two ABSs becomes spin-$z$ eigenstates   with two different eigenvalues, which have therefore   vanishing  matrix entries of the current. This shows the crucial role played by the magnetic doping in enabling the    ASQ manipulation.  Second, the   electric dipole transition amplitude is quite sensitive to the introduction of magnetic doping, since $g_{12}$ exhibits a steep increase even for a small reduction of transmission $T_\delta$ from 1. Thus, the   transition amplitude is a non negligible fraction of 1, even for a relatively weak impurity, i.e. for a transmission as high as $90 \%$.

\section{Manipulation of the ASQ}
\label{sec-4} 
\begin{figure}
\includegraphics[width=\linewidth]{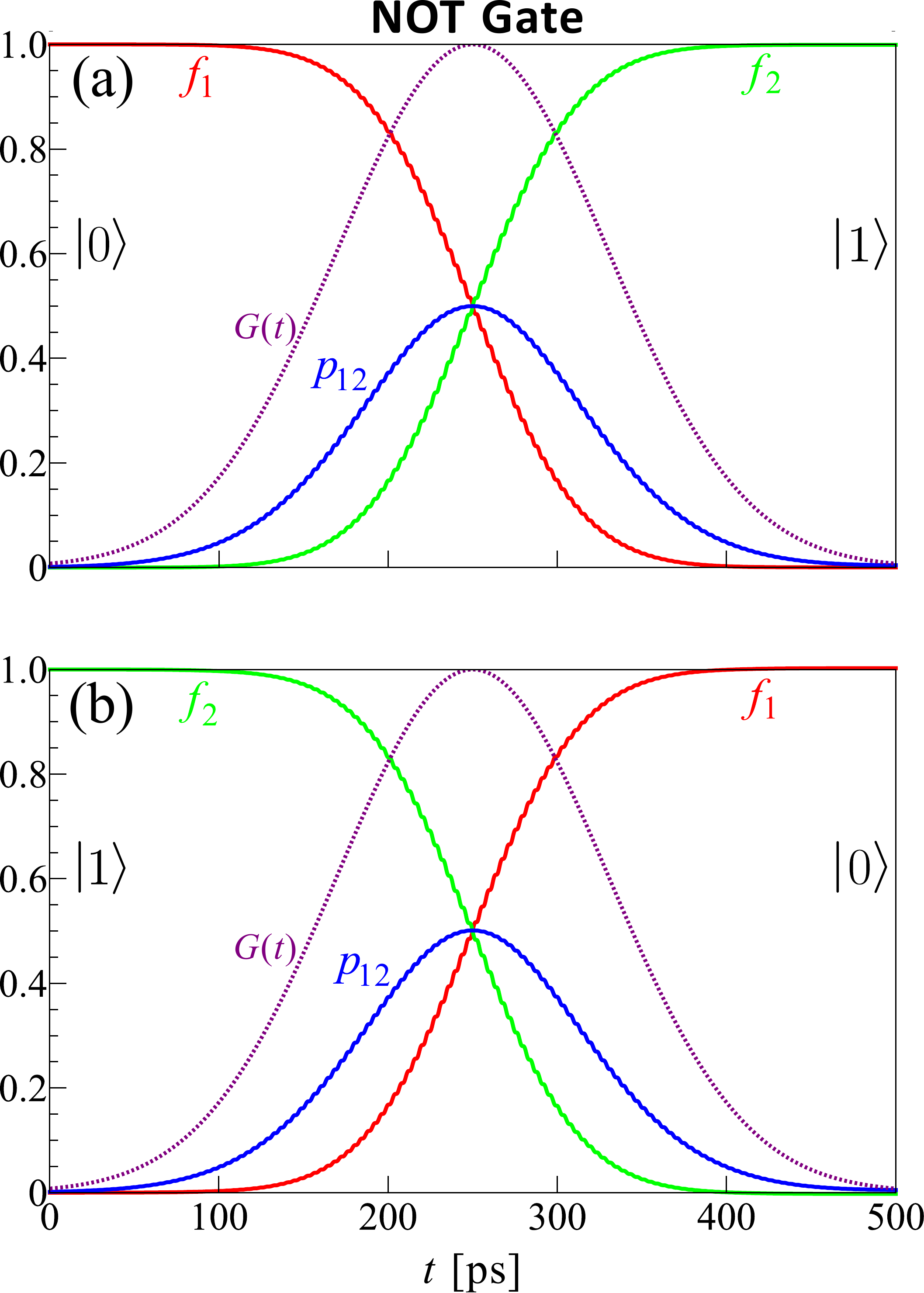}
\caption{\label{Fig6-NOT-gate}    Simulation of the implementation of a NOT gate in the QSHI-based ASQ.  A  magnetic $\delta$-impurity characterized by transmission coefficient $T_\delta=0.8$ is located at position $x_0=L/4$ inside the JJ weak link. The length parameter is $\lambda=2$, and the superconducting phase difference is $\phi=\pi/2$.  }
\end{figure}

In this section we demonstrate the possibility of realizing quantum logic gates on the QSHI-based ASQ,   by applying a properly tailored electromagnetic pulse. To this aim, we present a few prototypical simulated experiments, performed by a fixed-timestep solution of the Liouville-von Neumann equation~(\ref{LvN-Eq-Interaction}), where we have 
considered an electromagnetic excitation consisting of a sequence of independent Gaussian pulses,   each pulse being described by the vector potential
\begin{equation}\label{Gauss-A}
A(t) =   \frac{\mathcal{E}_0}{\omega} \, G(t)  \, \cos\left[\omega (t-\overline{t})+\varphi\right].
\end{equation}
Here, $\mathcal{E}_0$ denotes the magnitude of the electric field, $\omega$   the  pulse  frequency and $\varphi$ a phase, whereas
\begin{equation}\label{G(t)-def}
    G(t)=\exp\left[-\frac{(t-\overline{t})^2}{2 \tau^2}\right]  
\end{equation}
is a Gaussian time envelope  profile, centered around the time $\overline{t}$ and with a duration parameter $\tau$ (see e.g. the violet curve in Fig.\ref{Fig6-NOT-gate}). 
Moreover, we shall consider the  adiabatic excitation regime, i.e., $\omega  \tau  \gg 1$, where the  Gaussian envelope of each pulse contains a large number of oscillations, and the electric field corresponding to the vector potential in Eq.~(\ref{Gauss-A}) is well approximated  by a   Gaussian pulse  as well 
\begin{equation}\label{Gauss-E}
E(t) \simeq    \mathcal{E}_0 \, G(t) \, \sin\left[\omega  (t-\overline{t} )+\varphi \right]. 
\end{equation}

For definiteness, we consider a helical JJ realized with a HgTe/CdTe  QSHI proximized by Nb films. The Fermi velocity of the helical edge states is taken as $v_F=5 \times 10^{5} {\rm m/s}$~\cite{molenkamp-zhang_review}, and the induced superconducting pairing is assumed to be  $\Delta_0 = 1$\,meV. As an ASQ architecture, we start by considering a helical JJ with two ABSs,   obtained by taking the parameter $\lambda=2$, corresponding to a weak link length   $L \simeq 660 {\rm nm}$ [see  in Eq.(\ref{lambda-def})]. Moreover, we focus  on the regime of a localized magnetic impurity,  where the ABSs are  depicted by the red dashed curve of  Fig.~\ref{Fig2-ABS}. We have considered a Gaussian pulse with a  frequency   $\omega  = (E_2-E_1)/\hbar$ resonant with the ASQ, applied for a pulse duration $\tau  = 80$\,ps, around a  time  $\overline{t}  = 250$\,ps. 
We suppose that the ASQ can be  prepared  in the computational input states $\vert 0 \rangle$ or $\vert 1 \rangle$. The former corresponds  to a full occupation of the lower ABS ($f_1 = \rho^{ee}_{11} = 1$)  and an empty upper ABS ($f_2 = \rho^{ee}_{22} = 0$), the latter to the opposite situation ($f_1 =0,  f_2=1$), respectively. \\

A quantum logic gate transforms the input states $\vert 0 \rangle$ or $\vert 1 \rangle$ into specific output states, by performing an appropriate
 rotation by an angle $\theta$ in the  computational  basis space. 
Two noteworthy aspects are relevant to describe its realization. First,  the product $\mathcal{E}_0\, \tau$ of  the  amplitude  and the duration  of the  pulse must be suitably   calibrated to perform an on-demand rotation. For a two-level system  characterized by an    electric dipole  transition amplitude~$g_{12}$, the suitable value of $\mathcal{E}_0  \tau$ can be given the analytical expression~\cite{rossi-book} 
\begin{equation}\label{PulseAmplitude}
\mathcal{E}_0 \tau  = \frac{(E_2-E_1) \, \theta}{\sqrt{2\pi} \,{\rm e} v_F |g_{12}|  }
,
\end{equation}
provided  off-resonant coupling terms  are neglected.  
The second aspect to be emphasized is that, a priori,  the conditions to manipulate the ASQ depend  on the superconducting phase difference $\phi$ in a nontrivial way. Indeed, on the one hand, in order to maximize the interlevel energy splitting $E_2 - E_1$, it is advisable to operate at $\phi \neq 0$, preferably near $\phi \sim \pi$, as shown in Fig.~\ref{Fig2-ABS}. On the other hand, the  electric dipole  transition amplitude $|g_{12}|$ is minimal at such $\phi$-value, and maximal at $\phi=0$, as shown in Fig.~\ref{Fig4-g12}. Moreover, while for $\phi = 0$ the diagonal matrix elements $g_{11}$ and $g_{22}$ are absent,  they become significantly different from zero for increasing values of $\phi$, giving rise to energy renormalization effects during the ASQ   manipulation. 
We have performed our simulations at the trade-off value of   $\phi = \pi/2$, yielding $E_2-E_1 \simeq 0.52$\,meV, and $|g_{12}| \simeq 0.43$. However, as we shall discuss below, the results do not change significantly with other values of superconducting phase difference.\\

\begin{figure}
\includegraphics[width=\linewidth]{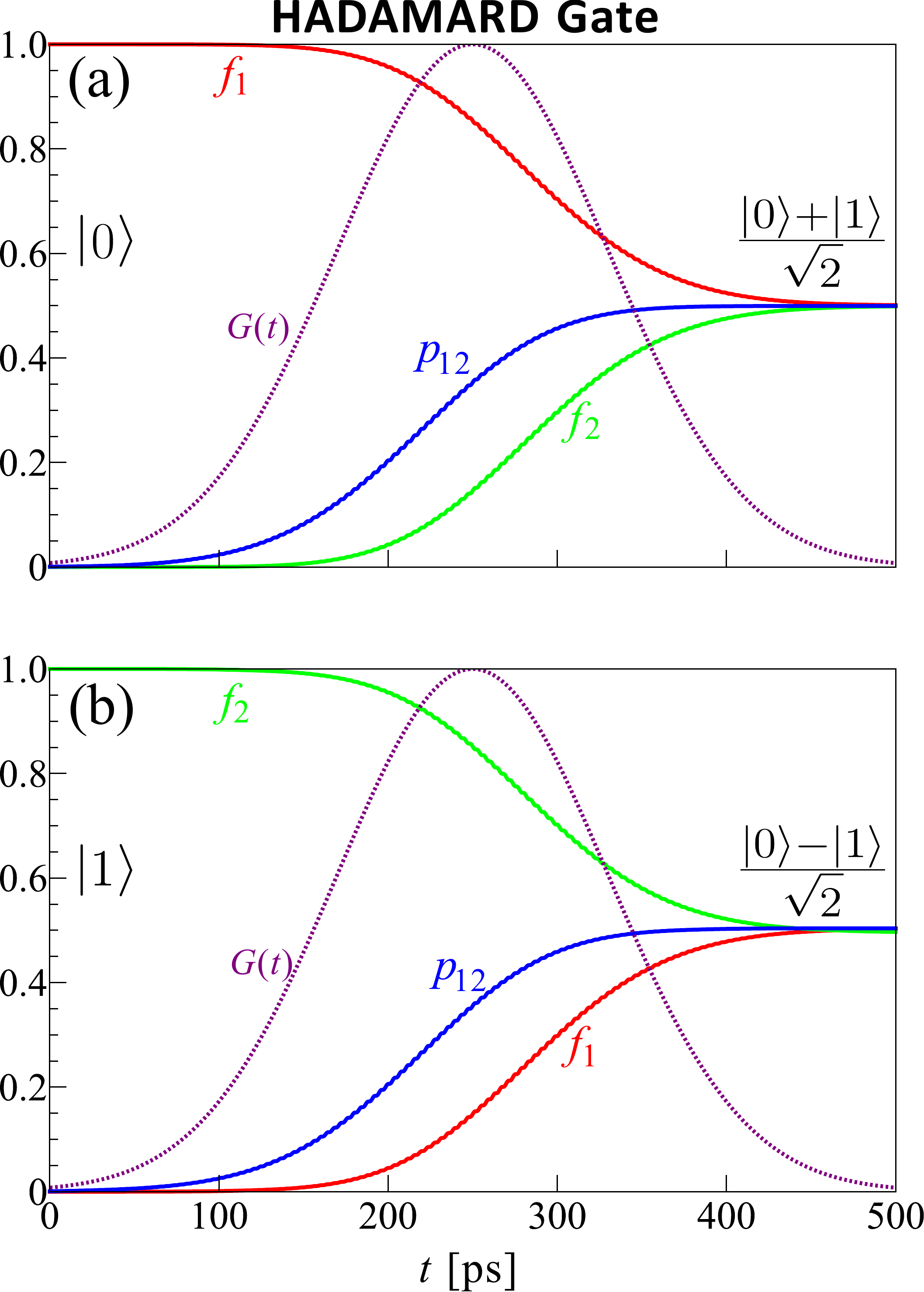}
\caption{\label{Fig7-HD-gate} Simulation of the Hadamard gate. The parameters are the same  as in Fig.~\ref{Fig6-NOT-gate}.}
\end{figure}

We start by describing the simulation of the quantum NOT gate,  which corresponds to a rotation by $\theta=\pi$ and converts $|0\rangle \rightarrow |1\rangle$  and $|1\rangle \rightarrow |0\rangle$.
The dynamical simulation is shown in Fig.~\ref{Fig6-NOT-gate}. Specifically,  panel (a) shows the  continuous transition of the  $|0\rangle$ input state into the $|1\rangle$ output state at the end of the pulse, while   panel (b) the reversed process $|1 \rangle \rightarrow |0 \rangle$. The red and green curves display the dynamical evolution of the ABS populations $f_1$ and $f_2$, respectively. The dotted magenta curve illustrates the time profile characterizing the applied pulse and described by the Gaussian  function $G(t)$ in Eq.(\ref{G(t)-def}).
Along with the change in the ABS populations, a nonvanishing value of the interlevel polarization  emerges during the two   excitation  processes. Indeed, the blue curves in Fig.\ref{Fig6-NOT-gate} illustrate the evolution of   its absolute value $p_{12} = \vert\rho^{ee}_{12}\vert$. As one can see, while $p_{12}$ is always vanishing before and after the pulse,  it  reaches a maximum value of $1/2$ exactly at the center of the Gaussian pulse, where also the two populations reach the very same value ($f_1 = f_2 = 1/2$), as dictated by the unitary dynamical evolution. \\

Let us now turn to the Hadamard gate, which plays a central role in many quantum information processing protocols~\cite{nielsen-chuang_book}. It corresponds to a rotation by an angle $\theta=\pi/2$, and  
operates as follows
\begin{eqnarray}
    \vert 0 \rangle & \rightarrow & \frac{1}{\sqrt{2}}\left(\vert 0 \rangle + \vert 1 \rangle\right),\label{HD-input-0} \\
    \vert 1 \rangle & \rightarrow & \frac{1}{\sqrt{2}}\left(\vert 0 \rangle - \vert 1 \rangle \right) ,\label{HD-input-1}
\end{eqnarray}
i.e. converting  the  ASQ input states into  the equally weighted quantum superpositions,
characterized by   $f_1 = f_2 =1/2$ and by $\rho^{ee}_{12} = \rho^{ee}_{21}=\pm 1/2$, respectively.  Its dynamical implementation is illustrated in   Fig.~\ref{Fig7-HD-gate}, where the two panels (a) and (b) describe the processes (\ref{HD-input-0}) and (\ref{HD-input-1}), respectively.  
Note that the corresponding population and polarization time profiles confirm that the ASQ state after the Hadamard gate processing  is equivalent to the ASQ state in the middle of a NOT gate (see Fig.\ref{Fig6-NOT-gate}).   \\

Although the above results have been obtained for the trade-off value $\phi=\pi/2$ of the superconducting phase difference, we have verified that the value of   $\phi$ has  in fact a negligible impact on the quality of the  quantum manipulation. For instance, at  
$\phi = 0$, the simulated experiments lead to results that are quite similar to the ones shown in   Figs.~\ref{Fig6-NOT-gate}  and~\ref{Fig7-HD-gate}, in spite of differences in the corresponding values of  interlevel energy splitting ($E_2-E_1 \simeq 0.25$\,meV) and  transition amplitude ($|g_{12}| \simeq 0.94$). The only minor feature that emerges for  $\phi \neq \pi/2$  is  the appearance of small and wiggly modulations of the two populations and of the interlevel polarization. The period of such small modulations corresponds to  a frequency $2\omega$ and is due to the off-resonant contributions that, while being often neglected in   models based on the rotating-wave approximation,   are fully taken into account by our approach.   
Apart from such a minor feature, the robustness of the result to the choice of $\phi$ proves   the accuracy of the simulated quantum gates. In turn, this  also shows that the presence of non-zero diagonal matrix elements $g_{11}$ and $g_{22}$ does not significantly affect the simulated experiments of Figs.~\ref{Fig6-NOT-gate} and ~\ref{Fig7-HD-gate}. This can be explained by observing that,  in the envisaged adiabatic-excitation regime  $\omega  \tau  \gg 1$, the Gaussian pulses contain a large number of oscillations, and therefore the energy renormalization effects mentioned in Sec.\ref{sec-3} effectively average out to zero.

Finally, a comment is in order about  the off-diagonal blocks $\tilde{\mathsf{g}}^{eh}_{\rm I}(t)$ and $\tilde{\mathsf{g}}^{he}_{\rm I}(t)$ in Eq.~(\ref{Hprime-Int-pic}),   which physically describe the simultaneous photoexcitation of two electrons from the superconducting condensate to the generic pair of ABS $i$ and $j$. In all the simulated experiments presented so far, which refer to the case of a JJ with two ABSs ($N = 2$) and involve an initial state with $f_1 + f_2 = 1$, the matrix entries $\tilde{g}_{ij}$  do not contribute at all to the time evolution of the density matrix, as   can also be proven by a few algebra steps.
However,   as we shall discuss in the next section, the $\tilde{g}_{ij}$ coefficients may play a crucial role when more ABSs are present.

\section{Discussion}
\label{sec-5} 
This section is devoted to discuss some aspects concerning the implementation in realistic systems, the preparation of the initial quantum state and the dissipation and decoherence effects.
 
\subsection{Possible experimental implementations}
\label{sec-5A}
The QSHI state has been experimentally observed in {\rm HgTe/CdTe}~\cite{zhang_2006,molenkamp_2007,brune_2010,molenkamp_2013} quantum wells, in {\rm InAs/GaSb}~\cite{knez_2011,knez_2015} bilayers, in  Bi  bilayers~\cite{murakami_2006,yazdani_2014,li_book_2019}, as well as in  ${\rm WTe}_2$ monolayers \cite{cobden_2017,shen_2017,jarillo-herrero_2018}. QSHI-based JJ have been realized using various superconductors. In particular,  Al contacts have been used in {\rm InAs/GaSb} implementations~\cite{kouwenhoven_2015}, ${\rm NbSe_2}$ contacts in ${\rm WTe_2}$ monolayers~\cite{hunt_2020}, while HgTe  has been proximized with Nb films~\cite{molenkamp_2016,molenkamp_2025}, Al contacts~\cite{yacoby_2013,molenkamp_natnanotech_2016} and MoRe electrodes~\cite{molenkamp_2024}.
Moreover, magnetic doping of QSHI has been recently achieved with  ferromagnetic Fe impurities  deposited on    Bi bilayers~\cite{yazdani_2020},  and with dilute Mn atoms in {\rm HgTe}-based implementations~\cite{molenkamp_2021,molenkamp_2024}.  

The simulations we have presented are inspired to a {\rm HgTe/CdTe}-based QSHI  proximized by Nb contacts, which seems to be most promising platform for the ASQ realizations,  since decoherence effects due to inelastic scattering mechanisms are expected to be reduced, as we  shall argue below in Sec.\ref{sec-5C}.
Moreover,
experimental evidence shows that in  HgTe/CdTe quantum wells edge channels   carry significantly more  current than what one could expect in a non-topological conductor~\cite{yacoby_2013,molenkamp_2013}, ruling out possible edge conduction of topologically trivial origin arising from unwanted edge doping   during fabrication or from band bending.

Three more aspects concerning the experimental implementations deserve a comment. First, in setups of proximized QSHI, the superconducting films typically cover both edges of the QSHI bar. However, because the two edges are usually separated  by a distance $W$ of several ${\rm \mu m}$, i.e. longer than the superconducting phase coherence length~\cite{yacoby_2013,molenkamp_2024}, the two weak links can be considered as independent. Thus,   our model, where only one edge has been considered,  reliably applies. 
A second comment is  about the possible   leakage of  quasi-particles  into the weak link (quasiparticle poisoning ~\cite{recher_2020,loss-klinovija_2022}). In QSHI-based JJ, it has been predicted that quasiparticle poisoning might induce a change in the Josephson current periodicity, when the superconducting phase $\phi$ varied~\cite{beenakker_2013,sassetti_2013,trauzettel_2014}. In an  ASQ, however,  one customarily operates around a fixed value of $\phi$, which sets the energy separation of the ASQ states and the radiation frequency to be used for the manipulation.  Still, quasiparticle poisoning should be limited experimentally, as it may alter the ABS occupancy.
Poisoning by quasiparticles originating from the superconductors can be reduced by modulating the thicknesses of the superconducting films, thereby trapping  quasiparticles sufficiently far away from the weak link~\cite{fatemi_prappl_2025}. 
Other sources of quasiparticle poisoning can be ascribed to an additional localized level present inside the weak link, which could effectively act as a quasiparticle `bath' for the ASQ, as has been observed in nanowire setups~\cite{fatemi_prappl_2025}. However, while in nanowires such bound states could emerge from a local confinement due to spin-orbit inhomogeneities,  this possibility is reduced for the helical states of a QSHI, whose  linear and helical spectrum hinders confinement.  
For these reasons, we expect that in the proposed setup the  quasiparticle poisoning effect can be mitigated, at least as much as can be done in current nanowire-based ASQ implementations.  The third aspect to be mentioned is that, different from conventional materials, the zero-energy crossing of the ABS spectrum at      $\phi=\pi$ shown in Fig.\ref{Fig2-ABS} is robust.
Indeed, due to the helical nature of the topological states, the number of degrees of freedom in a QSHI-based JJ is halved as compared to conventional JJ, and the ABS are non degenerate fermionic states. As a consequence,   the two energy levels 
$\pm E_1(\phi)$ crossing at zero-energy at    $\phi=\pi$ shown in Fig.\ref{Fig2-ABS} are related to two fermion parity sectors ($\gamma^\dagger_1\gamma^{}_1=0,1$, i.e. even/odd) of the excitation Hamiltonian $\mathcal{H}_1=E_1(\phi)  (\gamma^\dagger_1 \gamma^{}_1 -1/2  )$~\cite{fu-kane_2009,beenakker_2013}.
As a consequence, the zero-energy crossing is  topologically protected~\cite{yakovenko_2004,fu-kane_2009} and cannot even be removed by quasiparticle poisoning that switches the parity.

Finally, it is worth mentioning that the weak link of the JJ is typically embedded in a superconducting loop, which controls the superconducting phase difference through an external flux. In turn, the loop is also inductively coupled to a microwave resonator, usually an LC circuit, whose frequency depends on the occupancy of the ABSs. 
The manipulation of the ASQ can be induced  by applying either an AC signal through a gate (all-electric  control)   or by modulating the magnetic flux through  an AC current in a conductor coupled to the superconducting   loop~\cite{levy-yeyati_2017,pothier_2021,houzet-meyer_2024,pothier_2019,hays_2020,hays_2021,pita-vidal_2023,kouwenhoven_2023,pita-vidal_2024,pita-vidal_2025,fatemi_prappl_2025,fatemi_prl_2025}.
On the one hand, an all-electric control of the ASQ offers great advantages in terms of scalability, integration with classical control circuits on-chip, and overall simplicity in the setup structure. On the other hand, the timescale of electrical manipulation is the nanosecond, which may limit the number of operations that can be performed within the ASQ decoherence time. A possible alternative strategy could be an optical control performed with laser pulses. While this integrated technology is certainly more complex from the viewpoint of the setup architecture, 
in the case of spin qubits realized with semiconductor quantum dots, 
it was proven to enable the control in the picosecond or even femtosecond timescales, thereby increasing by orders of magnitude the number  of operations that can be performed within the decoherence time~\cite{awschalom_2001,awschalom_2008,yamamoto_2008,yamamoto_2013}.
\subsection{Preparation of the initial state}
\label{sec-5B}
\begin{figure}
\includegraphics[width=\linewidth]{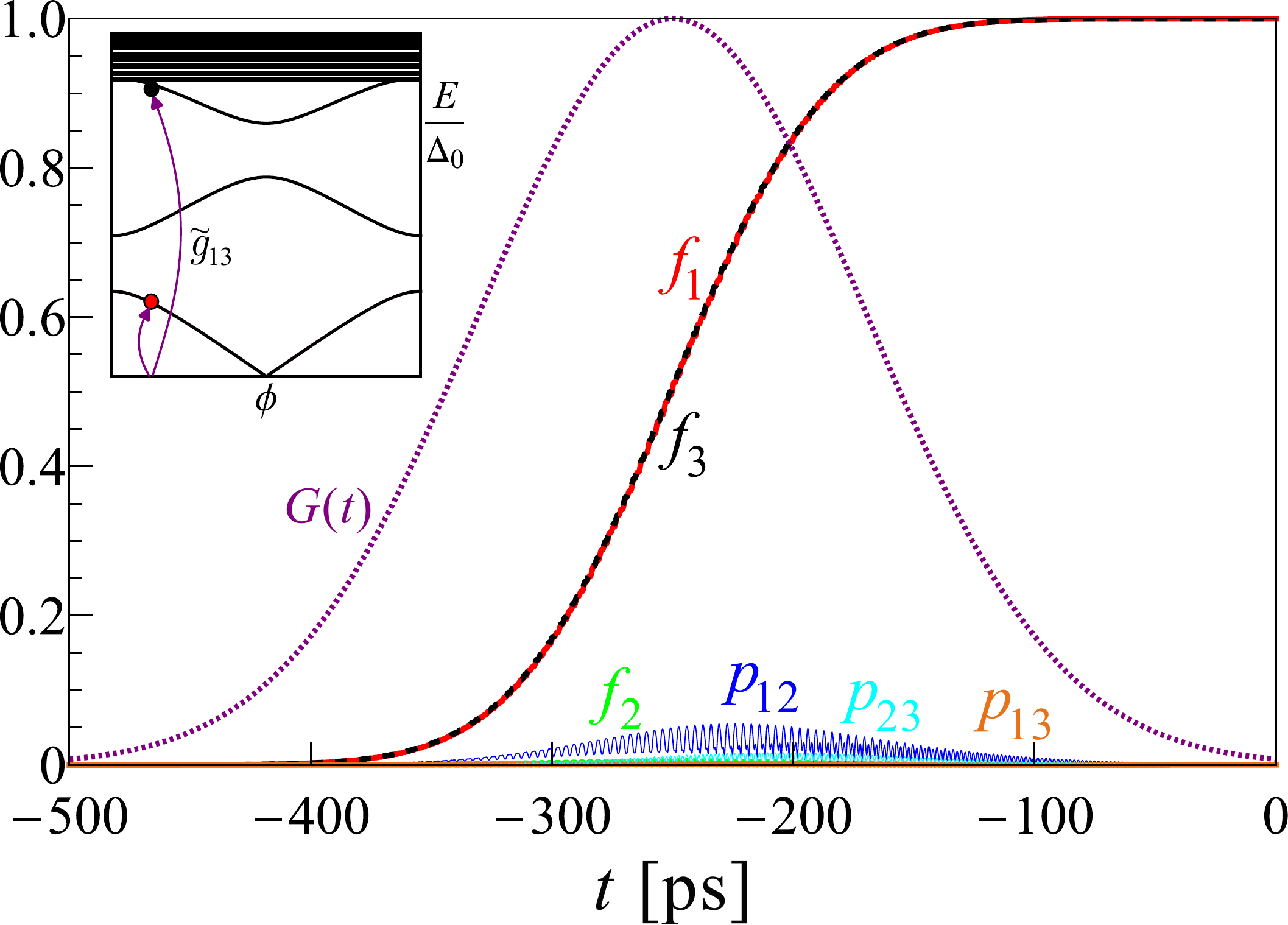}
\caption{\label{Fig8-SP} 
   Simulation of the initial state preparation (negative times) induced by an electromagnetic pulse  with $\overline{t}=-250 {\rm ps}$ and $\tau=80 {\rm ps}$. The preparation protocol is based on a JJ with  three ABSs, and a two quasi-particle photoexcitation from the condensate (see text), as depicted in the inset. The parameters of the weak link are $\lambda=3$ and $\phi=\pi/4$, while the magnetic $\delta$-impurity, located at position $x_0=L/4$ from the center of the weak link, has a transmission coefficient $T_\delta=0.8$.   } 
\end{figure}

In the previous Sec.~\ref{sec-4}, we have presented a few  simulations of quantum information processing assuming that the ASQ can be prepared   in an input state $\vert 0 \rangle$ or $\vert 1 \rangle$. Since in the low-temperature limit the equilibrium population of all the ABSs vanishes, it is crucial to identify a reliable protocol for the preparation of the initial computational state. 
As an illustrative example, let us discuss for instance   the preparation of the $|0\rangle$ state, characterized by an energy $E_1$.  The physical protocol that we aim to mimic, inspired by Ref.\cite{nazarov_2003},  consists in   photoexciting the condensate with a frequency $\Delta_0+E_1 < \hbar \omega < 2 \Delta_0$. In this way one quasi-particle is excited in the state with $E_1$, while the other one is promoted to the continuum spectrum, and plays no further role for the ASQ, for it leaves the weak link within a timescale of the order $\tau_e \sim L/v_F$. In order to simulate this protocol, in principle one would  need to include, together with the two discrete  ABS of the ASQ, also a huge number of closely spaced states with energies $E>\Delta_0$ describing the continuum. However, as far as the simulation of the initial state preparation is concerned, for practical computational purposes nothing changes to the ASQ preparation process if one considers a third discrete level, with an energy $E_3$ lower but very close to $\Delta_0$ (see inset of Fig.\ref{Fig8-SP}), that effectively plays the role of the continuum. This enables one  to mimic the role of the continuum spectrum, while retaining a limited number of states in the simulation.
Thus, starting from the low-temperature equilibrium state ($f_1 = f_2 = f_3 = 0$), one could fully promote   two electrons from the superconducting condensate to the lowest energy level $E_1$ and to the highest energy level $E_3$. This  type of process relies on the $\tilde{g}_{13}$ coefficient, and can be easily induced via a properly tailored electromagnetic pulse with frequency $\omega = (E_1+E_3)/\hbar$. Indeed, if the intermediate energy level $E_2$ is sufficiently separated from both level $E_1$ and   level $E_3$, the excitation pulse  will not change its low-temperature equilibrium population ($f_2 = 0$). As a result of such preliminary photoexcitation process, the three-level weak link is then promoted into the physical state $f_1 = 1$, $f_2 = 0$, and $f_3 = 1$, as sketched in the  inset of Fig.\ref{Fig8-SP}, and therefore the ASQ   is correctly prepared in its computational state $\vert 0 \rangle$ ($f_1 = 1$ and $f_2 = 0$).

The two requirements of (i) ABS  energy levels relatively far from each other and  (ii) an electric dipole transition amplitude $\tilde{g}_{13}$ significantly different from zero are fulfilled by choosing a superconducting phase difference $\phi = \pi/4$ in the three-ABS weak link described in the inset of Fig.~\ref{Fig8-SP}. 
Indeed, in this case one has $E_1 \simeq 0.25$\,meV, $E_2 \simeq 0.51$\,meV, and $E_3 \simeq 0.98$\,meV, as well as   $|g_{12}| \simeq 0.67$ and    $|\tilde{g}_{13}|\simeq 0.10$, which corresponds to an electron-pair generation energy $E_1+E_3 \simeq 1.24$\,meV and to an ASQ energy splitting $E_2-E_1 \simeq 0.26$\,meV. 
The result of simulated experiments  preparing the initial state $|0\rangle$ is shown in the negative time axis of Fig.~\ref{Fig8-SP}, where  the electron-pair photoexcitation  described above is realized via a properly tailored Gaussian pulse centered around the time  $\overline{t}=-250$\,ps and with a pulse duration of $80$\,ps. The accuracy of our simulated experiment fully confirms the reliability of the proposed computational-state preparation.

\subsection{Energy dissipation and decoherence phenomena}
\label{sec-5C}
As is well known~\cite{schoen-shnirman-review_2001,devoret_2004,wendin-shumeiko_2007,burkard-chirolli_2008,burkard_2023}, a major limitation of solid-state quantum information processing platforms is the presence of energy dissipation and decoherence (DD) phenomena.
Although helical states are topologically protected from elastic backscattering off time reversal symmetric perturbations, dissipation and decoherence can arise from inelastic processes. Here, we would like to analyze first the various possible sources of these effects in the proposed ASQ, and then provide an estimate of these effects in a worst case scenario.

\subsubsection{Inelastic scattering mechanisms}

{\it Hyperfine interaction.} We start with the hyperfine interaction, which couples   electron and   nuclear spins, and is considered  the main  origin of the relatively short decoherence time in the current ASQ implementations    with InAs nanowires~\cite{hays_2021,pita-vidal_2023,pita-vidal_2025,fatemi_prappl_2025,tahan_2025}. 
In QSHI realized with InAs/GaSb bilayers~\cite{knez_2011,knez_2015}, the hyperfine constant is of the same order of magnitude as nanowires~\cite{merkulov_2002,loss_2017,loss_2018}. However, in  HgTe/CdTe, due to the limited percentage of crystal atoms with a nonvanishing nuclear spin   (at most 17\%), and a smaller nuclear spin, the hyperfine coupling constant is smaller by more than one order of magnitude~\cite{platero_2012,rosenow_2013,loss_2018}. By Fermi golden rule, one  can therefore expect that spin relaxation times due to nuclear spin coupling are enhanced by roughly two orders of magnitude.

{\it Electron-phonon coupling.} Another possible inelastic scattering is the coupling to phonons. Although acoustic phonons  do not couple to the spin degree of freedom directly, their interplay with Rashba impurities present in the QSHI   might in principle break the topological protection of the edge states. However, this effect is known to have a quite negligible impact on the normal current through the helical edge state, at least to  leading order in the Rashba disorder strength~\cite{budich_2012,loss-hsu_2021}. We   expect that a similar situation occurs when the helical states are proximized by superconductors, and that phonon-induced dynamical spin-flip processes are not significant.

{\it Electron-electron interaction.} As is well known,   since in an ASQ charge  can fluctuate back and forth from the JJ weak link to the superconductors through Andreev reflections, Coulomb charging energy effects are strongly suppressed with respect to conventional quantum dots. Yet,  in the ASQ proposed here, where the  edge states of the QSHI  are one-dimensional, electron-electron interaction   can give rise to collective excitations, realizing a helical Luttinger liquid, characterized by a parameter $K \le 1$ that identifies the strength of the repulsive screened Coulomb interaction (with $K=1$ describing the non-interacting case). 
Spin decoherence caused by two-particle backscattering with spin-flip processes   can only occur at commensurate filling values (Umklapp processes), and is relevant only for $K<1/2$, i.e. for relatively strong interaction~\cite{zhang-bernevig_2006}.  
However, interaction can also interplay with   magnetic impurities\cite{loss-hsu_2021}. In the case of static impurities considered here,   single-particle  backscattering with spin-flip is relevant for $K<1$, thereby enhancing  the ABS spin texture modification, but with no impact on decoherence. In the presence of dynamical quantum impurities, such as Kondo impurities, the scenario depends on the interaction strength, on the possible impurity anisotropy  and on the temperature regime~\cite{zhang-bernevig_2006,maciejko_2009,tanaka_2011,johannesson_2012,altshuler_2013,cazalilla_2018,molenkamp_2021}. However, at temperatures below the Kondo temperature,  two-particle backscattering causing spin decoherence  are relevant only for very strong interaction ($K<1/4$). 

Thus, the decoherence effects heavily depend on the actual value of  the interaction strength $K$. Although experimental measurements of  $K$ are quite limited~\cite{du_2015}, various  theoretical studies have estimated that  in  QSHI realized with InAs/GaSb quantum wells the interaction is relatively strong ($K \simeq 0.4$), while in HgTe/CdTe based QSHI it is quite weak $K \lesssim 1$~\cite{maciejko_2009,teo-kane_2009,johannesson_2009,trauzettel-molenkamp_2019}.
Moreover, in the ASQ setup proposed here (see Fig.\ref{Fig1-setup}), where the helical Luttinger liquid is proximized by superconductors, further   aspects are noteworthy. 

On the one hand,   the problem of a  Luttinger liquid contacted to  superconductors is  computationally nontrivial, and analytical approaches are only viable in the case of tunnel  contacts~\cite{fazio-hekking_1995,fazio-hekking_1996} or in the regime of a very long junction $L \gg \xi_S$~\cite{maslov_1996}. 
On the other hand, the presence of two superconducting films deposited  on the QSHI is expected to entail two effects. First, an enhancement of the screening effects, thereby leading to a reduction of the effective electron-electron interaction and the related decoherence effects. Second, the finite length $L$ of the weak link ``cuts" the renormalization group flow at a finite Thouless energy scale $E_L=\hbar v_F/L$. Thus, at given values of interaction strength $K$ and weak link length $L$ characterizing a specific setup, the results obtained for the non-interacting case are expected to be quantitatively but not qualitatively modified by the interaction.
This expectation is confirmed, for instance, by the results known in the case of a strong magnetic impurity present inside a very long helical JJ, where the critical current has been shown to behave as $I_c \propto E_L^{4/K}$~\cite{sassetti_2013}. A similar modification occurs in the present case of an intermediate  length junction, where  the matrix entries $g_{ij}$ of the current operator (\ref{J-def}) are expected to depend in a $K$-dependent manner on an energy scale   that is a combination of the $E_L$ and $\Delta_0$. In HgTe/CdTe quantum wells, where $K \simeq 1$, the modifications with respect to the non-interacting case $K=1$ are expected to be negligible. \\

\subsubsection{Simulation of dissipation and decoherence effects in the worst case scenario}
A thorough analysis  taking into account microscopically all the above inelastic scattering mechanisms leading to DD processes in the proposed ASQ implementation is out of the scope of the present paper. Yet, we would like to  provide some quantitative information about the impact of these effects. As argued above, there are various motivations to expect that dissipation and decoherence effects in JJ based on QSHI should be suppressed as compared to current   nanowire implementations, particularly in the case of HgTe quantum wells proximized by Nb films, and our simulations in Sec.\ref{sec-4} were inspired by  such setup. Moreover, a quality factor of a quantum hardware is the ratio of  the  decoherence time to the typical operational time scale, rather than the decoherence time as an absolute time.

Thus, we have considered the ``worst case scenario'' where the  decoherence time in the proposed QSHI-based   ASQ is similar to the one of the current nanowire  implementations, and we have tested how many operations one could perform within such a timescale. 
To this purpose, we have adopted a generalized phenomenological $T_1$-$T_2$ model~\cite{devoret_2004,burkard_2023}, where $T_1$ and $T_2$ are the relaxation and decoherence timescales, respectively. Specifically, 
while the simulated experiments presented in Sec.\ref{sec-4} result from a fully coherent dynamics, here we have included DD effects  by adding to each of the four blocks of the Nambu density matrix $\boldsymbol{\rho}$ the following incoherent time evolution term
\begin{equation}\label{T1T2}
\frac{d\rho^{\alpha\beta }_{ij}}{dt}\biggl|_{\rm inc} = -\Gamma_{ij}\,\left(\rho^{\alpha\beta}_{ij}-(\rho_{eq})^{\alpha\beta}_{ij}\right),
\end{equation}
with 
\begin{equation}
\Gamma_{ij} = \frac{\delta_{ij}}{T_1} + \frac{1-\delta_{ij}}{T_2}
\ ,
\end{equation}
where  $\alpha,\beta=e,h$ and $\rho_{eq}$ denotes the (diagonal) equilibrium density matrix. The diagonal terms ($i = j$) relax to their equilibrium values within the energy-dissipation time   $T_1$, while the off-diagonal terms ($i \neq j$) decay to zero within the decoherence time  $T_2$. It is important to stress that, in order to preserve the positive-definite character of the density matrix, it is   necessary to choose $T_1 \geq 2 T_2$~\cite{rossi-book}.

\begin{figure}
\includegraphics[width=\linewidth]{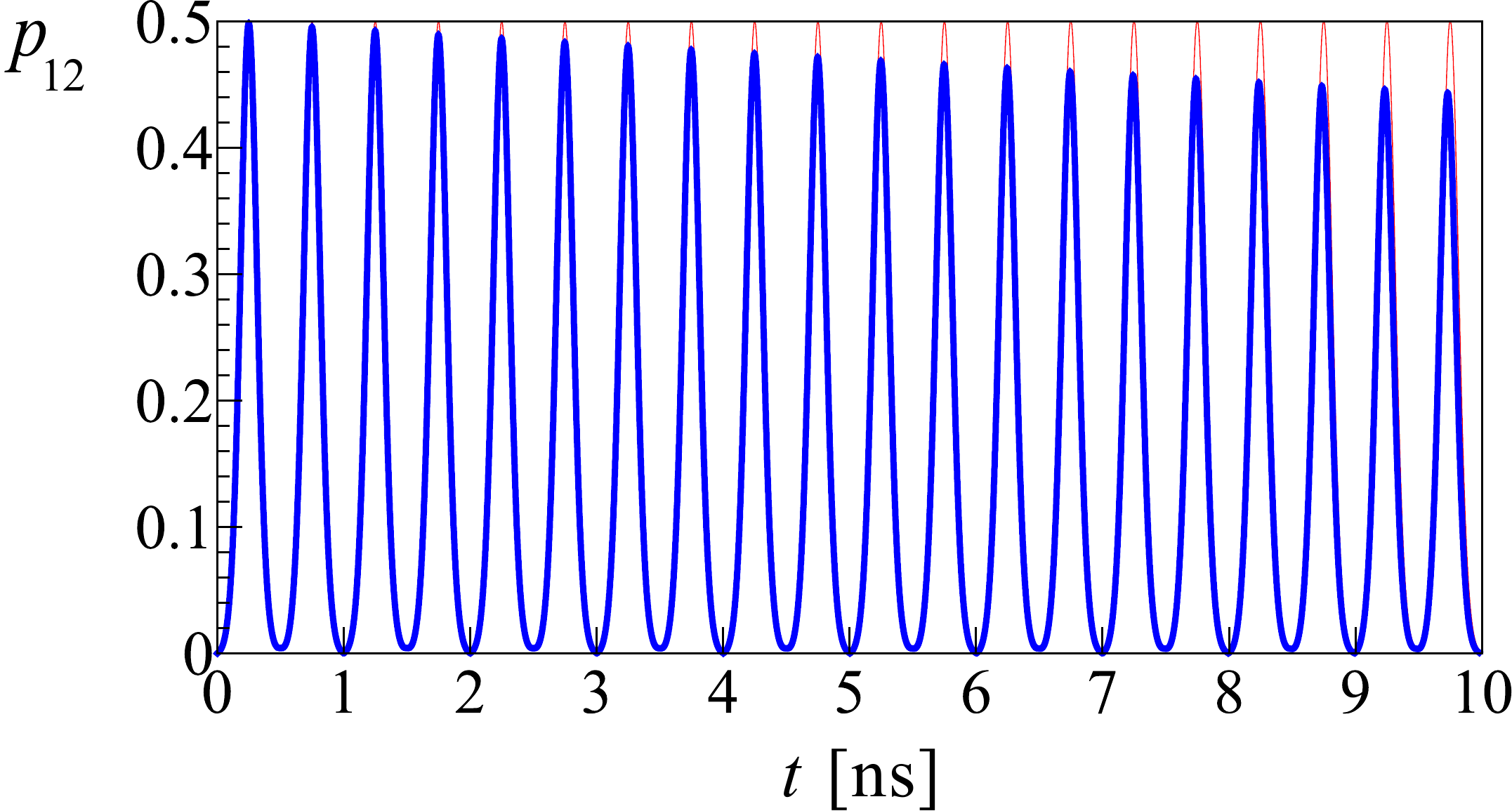}
\caption{\label{Fig9-decoherence} Simulation of a sequence of operations (20 NOT gates): time evolution of the polarization $p_{12}$.  The thin red line describes   the case of a unitary evolution, while the blue thick curve describes the behavior in  presence of dissipation and decoherence phenomena, accounted for via the phenomenological model in Eq.~(\ref{T1T2}) choosing $T_1 = 100$\,ns and $T_2 = 50$\,ns (see text). The JJ parameters are the same as in Fig.~\ref{Fig6-NOT-gate}. }
\end{figure} 

The simulated experiment presented in Fig.~\ref{Fig9-decoherence} corresponds to a sequence of 20 NOT gates and shows the time evolution of the interlevel polarization $p_{12}$, which is characterized by a sequence of maxima, one for each NOT gate (see Fig.~\ref{Fig6-NOT-gate}). In particular,  the thin red curve refers to the unitary evolution, while the blue thick curve describes the case where  DD phenomena are accounted for via the phenomenological model in Eq.~(\ref{T1T2}).  The decoherence time has been chosen as $T_2 = 50$\,ns, i.e. in agreement with the value characterizing  current ASQ implementations with nanowires \cite{hays_2021,pita-vidal_2023,pita-vidal_2024,fatemi_prappl_2025,fatemi_prl_2025},  while the relaxation time  has been set to  $T_1 = 100$\,ns, i.e.  the minimal compatible with the  constraint $T_1 \ge 2 T_2$ mentioned above. Here, the system as well as the simulation parameters are the same as in Fig.~\ref{Fig6-NOT-gate}. While in the fully coherent case (red curve) $p_{12}$ exhibits a periodic behavior, where all the polarization maxima reach the very same value ($p_{12} = 1/2$),  in the presence of DD phenomena (blue curve) a small decrease of the polarization maxima  occurs, which  is of the order of 10\% at the end of the simulation ($10$\,ns),  consistently with the decoherence time $T_2 = 50$\,ns used in the simulation. Our analysis shows that even in the worst case scenario, where the decoherence time is the same as in the current nanowire realizations, incoherent processes have a negligible impact on the realization of the   quantum gates based on the QSHI-JJ. Indeed,   the proposed ASQ manipulation allows one to realize tens of quantum operations  within the selected DD times.

\section{Conclusions}
\label{sec-6}
We have proposed an implementation of ASQs based on a JJ realized with the helical edge states of a QSHI, proximized by two superconducting films, as illustrated in Fig.\ref{Fig1-setup}. We have shown that the presence of magnetic doping alters the natural spin texture of the ABSs and leads to nonvanishing electric dipole transition amplitudes (see Figs.\ref{Fig4-g12}  and \ref{Fig5-g12-afo-Tdelta}), enabling the possibility to  manipulate the ASQ  with a radiation, without the need for an externally applied Zeeman field, and without invoking ancillary qubits. 
As illustrative examples, we have simulated the realization of   NOT and  Hadamard quantum gates  by   suitably tailored electromagnetic pulses applied to the QSHI-JJ, as shown in Figs.\ref{Fig6-NOT-gate}    and \ref{Fig7-HD-gate}, respectively. A possible strategy to prepare the initial state has also been outlined, and the impact of incoherent phenomena  in realistic realizations has been addressed.
In particular, the implementation with HgTe/CdTe QSHI proximized by Nb contacts  is suggested to be   promising, since decoherence effects   due to   Coulomb   and hyperfine interaction  are weak. 
Thus, the proposed realization of ASQ  might foster interdisciplinary research   across  topological materials and quantum information.

\section{Acknowledgements}
The authors greatly acknowledge interesting and inspiring discussions with F. Taddei, A. Braggio, A. Richaud and A.Crippa. Financial support from the  TOPMASQ ("Topological material platform for the implementation of Andreev spin qubit") project,   a Cascade call project (CUP E13C24001560001) financed by the “National Quantum Science \& Technology Institute”, PE00000023 (Next Generation EU) is also acknowledged.

\appendix
\section{ABS wavefunctions}
\label{AppA}
\begin{figure}
\includegraphics[width=\linewidth]{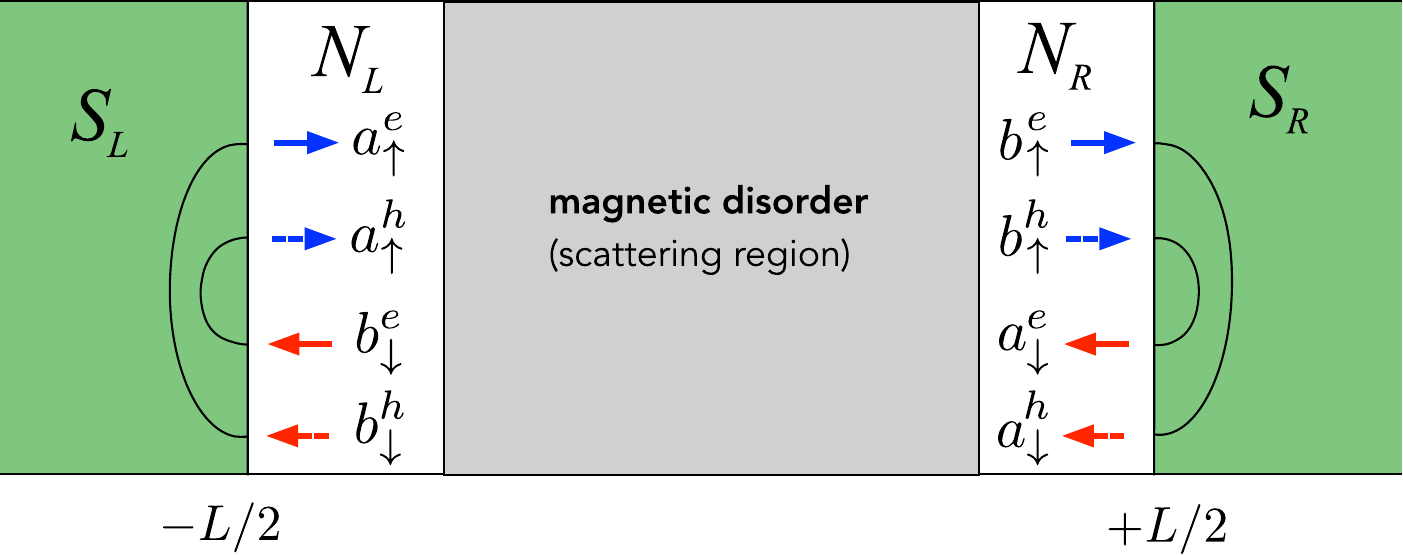}
\caption{\label{Fig-AppA} Scheme of the JJ:  Andreev scattering occurs at the interfaces $x_L=-L/2$ and $x_R=+L/2$, whereas the magnetic disorder inside the weak link induces normal scattering. The $a$ symbols denote the amplitudes of states that are travelling from the S interfaces towards the magnetic disorder regions, while the $b$ symbols the amplitudes of states travelling from the disordered region to the interfaces.}
\end{figure}

In this Appendix, we provide some details about the calculations of the ABS wavefunctions, which are needed to evaluate the  electric dipole transition amplitudes  in Eqs.(\ref{g_ij-def}) and (\ref{gtilde_ij-def}) shown in Sec.\ref{sec-3}. Following the scheme of Ref.\cite{beenakker_1991}, we schematize the JJ as depicted in Fig.\ref{Fig-AppA}, where the weak link,  sandwiched between the two superconductors $S_L$ and $S_R$, can be regarded   as a central scattering region characterized by the magnetic disorder, separated by ideal ballistic buffer normal regions $N_L$ and $N_R$. While Andreev scattering occurs at the interfaces $x_L=-L/2$ and $x_R=+L/2$ of the weak link with the superconductors, normal scattering occurs because of magnetic disorder.

An  eigenfunction $\Phi_{E}(x)$ of the BdG Hamiltonian~(\ref{H-BdG}) at   energy $E$
\begin{equation}\label{Phi-NR-side}
 \Phi_{E}(x)  =\begin{pmatrix}
u_{E  \uparrow}(x)  \\
u_{E \downarrow}(x)  \\
v_{E \downarrow}(x)   \\
v_{E \uparrow}(x)
\end{pmatrix}
\end{equation}
can be constructed by combining the solutions in the various JJ regions depicted in Fig.\ref{Fig-AppA}. Specifically, because we are interested in the ABSs, i.e. in the subgap spectrum $0 \le E \le \Delta_0$, the expression of the wavefunction in the superconductors and in the buffer regions reads as follows. 

\noindent {\it  Solutions in left superconductor $S_L$ (subgap)} 
\begin{eqnarray}\label{Phi-SL-side-pre}
\left. \Phi_{E}(x) \right|_{S_L} 
 = \sqrt{\frac{\Delta_0}{2E}} \, \begin{pmatrix} f_{-} e^{ -\frac{i}{2} \arccos\frac{E}{\Delta_0}}\, e^{i q_{E}^{-}x} \\ g_{+} \, e^{ \frac{i}{2} \arccos\frac{E}{\Delta_0}}\,  \, e^{-i q_{E}^{+}x} \\
  f_{-}\, e^{+ \frac{i}{2} \arccos\frac{E}{\Delta_0}} \, \, e^{i q_{E}^{-}x} \, e^{-i \phi_L} \\ g_{+} \, e^{ -\frac{i}{2} \arccos\frac{E}{\Delta_0}}\, e^{-i q_{E}^{+}x}\, e^{-i \phi_L}
\end{pmatrix}.\hspace{0.4cm}
\end{eqnarray}
\noindent {\it Solutions in the left buffer region $N_L$}
\begin{equation}\label{Phi-NL-side-bis}
\left. \Phi_{E}(x) \right|_{N_L}=      \begin{pmatrix}
a^e_{\uparrow} \, e^{i k^e_E x}\\
b^e_{\downarrow} \, e^{-i k^e_E x}\\
b^h_{\downarrow} \, e^{i k^h_E x}\\
a^h_{\uparrow}\, e^{-i k^h_E x}
\end{pmatrix}.
\end{equation}
\noindent {\it Solutions in the right buffer region $N_R$}
\begin{equation}\label{Phi-NR-side-bis}
\left. \Phi_{E}(x) \right|_{N_R}=     \begin{pmatrix}
b^e_{\uparrow} \, e^{i k^e_E x}\\
a^e_{\downarrow} \, e^{-i k^e_E x}\\
a^h_{\downarrow} \, e^{i k^h_E x}\\
b^h_{\uparrow}\, e^{-i k^h_E x}
\end{pmatrix}.
\end{equation}
\noindent  {\it Solution in the right superconductor $S_R$ (subgap)}
\begin{eqnarray}\label{Phi-SR-side-pre}
\left. \Phi_{E}(x) \right|_{S_R}  
=\displaystyle \sqrt{\frac{\Delta_0}{2E}} \begin{pmatrix} f_{+} e^{ +\frac{i}{2} \arccos\frac{E}{\Delta_0}}\, e^{i q_{E}^{+}x} \\ g_{-}\, e^{- \frac{i}{2} \arccos\frac{E}{\Delta_0}} \, e^{-i q_{E}^{-}x}\\
f_{+}  e^{- \frac{i}{2} \arccos\frac{E}{\Delta_0}} \, e^{i q_{E}^{+}x} \, e^{-i \phi_R}\\ g_{-}\,  e^{+\frac{i}{2} \arccos\frac{E}{\Delta_0}} \, e^{-i q_{E}^{-}x}\, e^{-i \phi_R} \hspace{0.3cm}
\end{pmatrix}  .\hspace{0.4cm}
\end{eqnarray}
In the formulas here above, the symbols $a$, $b$, $f$ and $g$ denote complex amplitudes,  where their energy label $E$ has been dropped to make the notation lighter. Moreover,   $\phi_L=\phi/2$ and $\phi_R=-\phi/2$ are the superconducting phases in left and right superconductors, while 
\begin{equation}\label{k^{e/h}_E}
    k^{e/h}_E = \frac{\mu \pm E}{\hbar v_F}
\end{equation}
denote the (real) wavevectors characterizing the normal buffer regions, and
\begin{eqnarray}
q_{E}^{\pm} =\frac{1}{\hbar v_F }\left(\mu \pm i\sqrt{\Delta_0^2-E^2 }\right),   \hspace{0.5cm}  0< E <\Delta_0,\,\,\,
\end{eqnarray}
the (complex) wavevectors characterizing the decaying states in the superconducting regions $S_{L/R}$ in the sub-gap.
 
Imposing the continuity of the wavefunction at the left and right interfaces $x=x_L=-L/2$ and $x=x_R=+L/2$
\begin{eqnarray}
\left. \Phi_{E}(x_L) \right|_{S_L} &=& 
\left. \Phi_{E}(x_L) \right|_{N_L},
\\
\left. \Phi_{E}(x_R) \right|_{S_R}&=&
\left. \Phi_{E}(x_R) \right|_{N_R} ,
\end{eqnarray}
one obtains the coefficients characterizing the superconductor wavefunctions
\begin{eqnarray}
   f_{-}\sqrt{\frac{\Delta_0}{2E}}&=& a^e_\uparrow  \, e^{ \frac{i}{2} \arccos\frac{E}{\Delta_0}}    \, e^{-i q^{-}_E x_L}\, e^{i k^e_E x_L},  \label{f_{-}-result}
\\
   f_{+}\sqrt{\frac{\Delta_0}{2E}}&=& a^h_\downarrow  e^{ \frac{i}{2} \arccos\frac{E}{\Delta_0}}   \, e^{-i q^{+}_E x_R}\, e^{i k^h_E x_R} \, e^{i \phi_R} ,\label{f_{+}-result}\\ 
   g_{-}\sqrt{\frac{\Delta_0}{2E}}&=& a^e_\downarrow \, e^{ \frac{i}{2} \arccos\frac{E}{\Delta_0}}   \,   e^{i q^{-}_E x_R}\, e^{-i k^e_E x_R},  \label{g_{-}-result}\\
      g_{+}\sqrt{\frac{\Delta_0}{2E}}&=& a^h_\uparrow  \, e^{  \frac{i}{2} \arccos\frac{E}{\Delta_0}}   \,   e^{i q^{+}_E x_L}\, e^{-i k^h_E x_L} \, e^{i \phi_L} , \,\,\, \,
 \label{g_{+}-result}
\end{eqnarray}
as well as the Andreev reflection relations
\begin{eqnarray} 
  b^e_\uparrow &=& a^h_\downarrow \, e^{  i \arccos\frac{E}{\Delta_0}}   \, e^{i \phi_R}\, e^{- i \frac{2E x_R}{\hbar v_F} },
\label{b^e_up-Andreev-scattering} \\
  b^h_\uparrow &=& a^e_\downarrow \, e^{ i \arccos\frac{E}{\Delta_0}}   \, e^{-i \phi_R}\, e^{- i \frac{2E x_R}{\hbar v_F} },
\label{b^h_up-Andreev-scattering}\\  
b^h_\downarrow &=&a^e_\uparrow \, e^{  i \arccos\frac{E}{\Delta_0}}   \, e^{-i \phi_L}\, e^{  i \frac{2E x_L}{\hbar v_F} }, \label{b^h_dn-Andreev-scattering} \\
b^e_\downarrow &=& a^h_\uparrow \, e^{  i \arccos\frac{E}{\Delta_0}}  e^{i \phi_L} \, e^{  i \frac{2E x_L}{\hbar v_F} }. \label{b^e_dn-Andreev-scattering}
\end{eqnarray} 
At the same time, the magnetic disorder present in the JJ weak link induces normal scattering processes. In the   electron sector, the scattering matrix $S_e$, which can be expressed as in Eq.(\ref{S0-e-gen}), connects  
\begin{equation}
\left(\begin{array}{l}
{b}^{e}_{\downarrow}    \\ \\ {b}^{e}_{\uparrow} \end{array} \right) = 
 \, 
 S_e(E) 
\left(\begin{array}{l}    {a}^{e}_{\uparrow} \\ \\ {a}^{e}_{\downarrow}   \end{array} \right)  ,   \label{Se-def}
\end{equation} 
whereas in the hole sector the Scattering Matrix $S_{h}(E)$, defined  through
\begin{equation}
\left(\begin{array}{l}
{b}^{h}_{\downarrow} \\   \\   {b}^{h}_{\uparrow} \end{array} \right) =  S_{h}(E) 
\left(\begin{array}{l}    {a}^{h}_{\uparrow} \\ \\{a}^{h}_{\downarrow}   \end{array} \right)   ,  \label{Sh-def}
\end{equation} 
can be obtained from $S_e$  through the relation (\ref{Sh-Se}). 
Thus,  in the buffer  regions $N_L$ and $N_R$ the outgoing amplitudes can be expressed in terms of the incoming ones as
\begin{eqnarray}
b^{e}_{\uparrow} &=& t_e \, a^{e}_{\uparrow} +r^\prime_e  a^{e}_{\downarrow}, \label{b^e_up-normal-scattering}\\  
b^{e}_{\downarrow} &=& r_e \, a^{e}_{\uparrow} +t^\prime_e  a^{e}_{\downarrow},\label{b^e_dn-normal-scattering} \\
b^{h}_{\uparrow} &=& t_h \, a^{h}_{\uparrow} +r^\prime_h  a^{h}_{\downarrow}, \label{b^h_up-normal-scattering}\\
b^{h}_{\downarrow} &=& r_h \, a^{h}_{\uparrow} +t^\prime_h  a^{h}_{\downarrow}. \label{b^h_dn-normal-scattering}
\end{eqnarray} 
Equating  Eq.(\ref{b^e_up-Andreev-scattering}) to (\ref{b^e_up-normal-scattering}), and  (\ref{b^h_up-Andreev-scattering}) to Eq.(\ref{b^h_up-normal-scattering}), one obtains
\begin{equation}
\begin{pmatrix}
    a^e_\uparrow \\ \\a^h_\uparrow
\end{pmatrix}= A_R \begin{pmatrix}
    a^e_\downarrow \\ \\ a^h_\downarrow
\end{pmatrix},
\end{equation}
where
\begin{equation}
 A_R=   \begin{pmatrix}
    \displaystyle -\frac{r^\prime_e}{t^{}_e} & \displaystyle \frac{1}{\alpha \,t^{}_e}e^{i \phi_R} e^{-i \frac{2  E x_R}{\hbar v_F}} \\  
    \displaystyle \frac{1}{\alpha\, t^{}_h}e^{-i \phi_R} e^{-i \frac{2  E x_R}{\hbar v_F}} & \displaystyle  -\frac{r^\prime_h}{t^{}_h}
\end{pmatrix} \label{AR-def}
\end{equation}
and 
\begin{equation}
\alpha(E)= \displaystyle  e^{-i \arccos\frac{E}{\Delta_0}}  . \label{alpha-def}
\end{equation}
Similarly, equating  Eq. (\ref{b^e_dn-Andreev-scattering}) to (\ref{b^e_dn-normal-scattering}), and Eq. (\ref{b^h_dn-Andreev-scattering}) to (\ref{b^h_dn-normal-scattering}), one retrieves
\begin{equation}\label{Eq-with-AL}
\begin{pmatrix}
    a^e_\downarrow \\ \\a^h_\downarrow
\end{pmatrix}= A_L \begin{pmatrix}
    a^e_\uparrow \\ \\ a^h_\uparrow
\end{pmatrix},
\end{equation}
where
\begin{equation}
A_L=\begin{pmatrix}
    \displaystyle -\frac{r^{}_e}{t^{\prime}_e} & \displaystyle \frac{1}{\alpha \,t^{\prime}_e}e^{i \phi_L} e^{ i \frac{2  E x_L}{\hbar v_F}} \\  
    \displaystyle \frac{1}{\alpha\, t^{\prime}_h}e^{-i \phi_L} e^{i \frac{2  E x_L}{\hbar v_F}} & \displaystyle  -\frac{r^{}_h}{t^{\prime}_h}
\end{pmatrix} . \label{AL-def}
\end{equation}
Combining Eqs.(\ref{AR-def}) and (\ref{AL-def}), one obtains
\begin{equation}\label{M2-pre}
    \left( \mathbb{I}_2-A_R\, A_L\right)\begin{pmatrix}
    a^e_\uparrow \\   \\ a^h_\uparrow
\end{pmatrix}=\begin{pmatrix}
    0 \\  \\ 0
\end{pmatrix},
\end{equation}
which has  two important implications. \\

On the one hand, Eq.(\ref{M2-pre}) determines    the equation $
{\rm det}\left( \mathbb{I}_2-A_L\, A_R\right)=0
$ of the 
  discrete energy levels $E_j$ characterizing the ABS wavefunctions $\Phi_E(x)$. Explicitly, using Eqs.(\ref{AR-def}) and (\ref{AL-def}), such an equation can be written as
\begin{eqnarray}
\lefteqn{\alpha^{-2} e^{-2 i \frac{E}{E_L}}\,+ \alpha^{2} e^{2 i \frac{E}{E_L}}  {\rm det}S^e(E)\,{\rm det}S^h(E) } & &  \nonumber \\
& & \,\,\, - \left(r^{\prime}_e r^{\prime}_h+r^{}_e r^{}_h  \right)-e^{i \phi}t^{}_e t^\prime_h - e^{-i \phi}t^{}_h t^\prime_e\,=0, \,\,\, \label{ABS-eq-pre}
\end{eqnarray}
where $\phi=\phi_L - \phi_R$ and, from Eq.(\ref{Sh-Se}), 
one has
\begin{eqnarray} 
r_h(E) &=& -r^*_e(-E), \\
t_h(E) &=& +t^*_e(-E), \\
t^\prime_h(E) &=& +{t^\prime_e}^*(-E),  \\
r^\prime_h(E) &=& -{r^\prime_e}^*(-E).
\end{eqnarray} 
Exploiting the second expression in Eq.(\ref{S0-e-gen}), one has ${\rm det}S^e(E) =-e^{2 i \Gamma_m(E)}$ and ${\rm det}S^h(E)=-e^{-2 i \Gamma_m(-E)}$. Moreover, using the definitions in Eq.(\ref{PhiA-GammaA-chiS-bis}), the ABS energy level equation (\ref{ABS-eq-pre}) reduces to Eq.(\ref{ABS-gen-fin}) given in the Main Text, recovering the result of Ref.\cite{dolcini_2014}.\\

On the other hand, Eq.(\ref{M2-pre}) also implies
\begin{equation}
 \displaystyle  a^h_\uparrow =\displaystyle \frac{ \alpha \, e^{i \frac{E}{E_L}} t^\prime_h\,  {\rm det}S_e  +\alpha^{-1} e^{-  i \frac{E}{E_L}}   \, e^{-i \phi}\, t^\prime_e }{  e^{-i \frac{\phi}{2}}  r^{}_h t^\prime_e + e^{i \frac{\phi}{2}}  r^{\prime}_e  t^\prime_h   } \,  \, a^e_\uparrow , \label{a^h_uparrow-ABS}  
\end{equation}
which can be substituted   into   the right-hand side of Eq.(\ref{Eq-with-AL}), leading to obtain through a few algebra steps   
\begin{equation}
a^e_\downarrow = \frac{\alpha^{-2} e^{-  2 i \frac{E}{E_L}}e^{-i \frac{\phi}{2}} -\left(e^{-i \frac{\phi}{2}} r^{}_e r^{}_h+e^{i \frac{\phi}{2}} t^{}_e t^\prime_h\right) }{e^{-i \frac{\phi}{2}}  r^{}_h t^\prime_e + e^{i \frac{\phi}{2}}  r^{\prime}_e  t^\prime_h} \, a^e_\uparrow  \label{a^e_downarrow-ABS} 
\end{equation}
and
\begin{equation}
 a^h_\downarrow = \frac{   \alpha^{-1} e^{-  i \frac{E}{E_L}} r^\prime_e - \alpha  \, e^{   i \frac{E}{E_L}} r^{}_h\, {\rm det}S_e  }{  e^{-i \frac{\phi}{2}}  r^{}_h t^\prime_e + e^{i \frac{\phi}{2}}  r^{\prime}_e  t^\prime_h} \, a^e_\uparrow.  \label{a^h_downarrow-ABS}
\end{equation}
Moreover, inserting Eqs.(\ref{a^h_uparrow-ABS})-(\ref{a^e_downarrow-ABS})-(\ref{a^h_downarrow-ABS}) into Eqs.(\ref{b^e_up-Andreev-scattering}), (\ref{b^h_up-Andreev-scattering}), (\ref{b^h_dn-Andreev-scattering}) and (\ref{b^e_dn-Andreev-scattering}), one finds
\begin{eqnarray}
b^e_\uparrow &=& \,  \frac{   \alpha^{-2} e^{- 2 i \frac{E}{E_L}} r^\prime_e -   r^{}_h\, {\rm det}S_e  }{r^{}_h t^\prime_e+e^{i \phi} r^\prime_e t^\prime_h} \, a^e_\uparrow ,\label{b^e_uparrow-ABS}\\
b^h_\uparrow &=& \frac{    \, e^{+i\frac{\phi}{2}}\, e^{- i \frac{E}{E_L} }}{\alpha} \times \label{b^h_uparrow-ABS}\\
& & \times \frac{\alpha^{-2} e^{- 2 i \frac{E}{E_L}} -\left(r^{}_e r^{}_h+e^{i \phi} t^{}_e t^\prime_h\right) }{r^{}_h t^\prime_e+e^{i \phi} r^\prime_e t^\prime_h} \, a^e_\uparrow \nonumber ,\\
b^h_\downarrow &=&  \, \alpha^{-1}   \, e^{-i\frac{\phi}{2}}\, e^{- i \frac{E}{E_L} } \, a^e_\uparrow ,\label{b^h_dnarrow-ABS}\\
b^e_\downarrow &=& 
    \displaystyle    \, \,   \frac{   t^\prime_h\, e^{  i \phi }\,  {\rm det}S_e  +\alpha^{-2} e^{- 2 i \frac{E}{E_L}}   \,  \, t^\prime_e }{ r^{}_h t^\prime_e + e^{i  \phi}   r^{\prime}_e  t^\prime_h   } \,  \, a^e_\uparrow.\label{b^e_dnarrow-ABS}
\end{eqnarray}
Equations (\ref{a^h_uparrow-ABS}), (\ref{a^e_downarrow-ABS}), (\ref{a^h_downarrow-ABS}), (\ref{b^e_uparrow-ABS}), (\ref{b^h_uparrow-ABS}), (\ref{b^h_dnarrow-ABS}) and (\ref{b^e_dnarrow-ABS}) express the 7 coefficients $a^h_\uparrow$, $a^e_\downarrow$,  $a^h_\downarrow$, $b^e_\uparrow$, $b^h_\uparrow$,  $b^h_\downarrow$ and $b^e_\downarrow$
as a function of the coefficient  $a^e_\uparrow$. Thus, inserting Eqs.(\ref{a^h_uparrow-ABS}), (\ref{a^e_downarrow-ABS}), and (\ref{a^h_downarrow-ABS}) into Eqs.(\ref{f_{+}-result}), (\ref{g_{-}-result}), (\ref{g_{+}-result}), the entire wavefunction in the superconductors and in the buffer regions can be expressed only in terms of the amplitude $a^e_\uparrow$ and the energy eigenvalue $E_j$ found by solving Eq.(\ref{ABS-gen-fin}).
Moreover, also the wavefunction inside the magnetic domain  can   be expressed in terms of $a^e_\uparrow$, through relations that depend on the specific envisaged scatterer. Here below, we show how this is done in the  illustrative example  of a magnetic barrier. Thus,  the entire wavefunction $\Phi_E(x)$ can be expressed in terms of $a^e_\uparrow$ only, which in turn is determined through the normalization condition $\int_{-\infty}^{+\infty} \Phi^\dagger_E(x) \Phi_E(x)\, dx=1$.

\subsection{Scattering matrix for a finite barrier}
Because the magnetic disorder does not couple electron and hole sectors, one can determine the electron and hole wavefunction separately. Let us consider the case of a  magnetic barrier 
\begin{equation}
    \mathbf{m}(x)= \left\{ \begin{array}{lcl} 0, & & x < x_1, \\ (m_\perp \cos\phi_{\perp}, m_\perp \sin\phi_{\perp}, m_z), & & x_1 <x < x_2, \\
    0, & & x < x_2,
    \end{array} \right.
\end{equation}
where $x_1>-L/2$ and $x_2 < L/2$ are the magnetic domain boundaries within the weak link. This case generalizes Eq.(\ref{barrier}) by including  also a $m_z$ component parallel to the natural quantization axis of the helical states, as well as a generic orientation $\phi_\perp$ within the spin $x$-$y$ plane of the magnetization.

We focus  on the electron sector, while a similar calculation can be carried out for the hole sector.  
The wavefunction in the clean buffer regions $N_L$ and $N_R$ reads
\begin{equation}\label{u-NL}
\begin{pmatrix}u_{E\uparrow}(x) \\ \\ u_{E\downarrow}(x) \end{pmatrix} =   \begin{pmatrix}\, {a}^{e}_{\uparrow}\,e^{+i k^{e}_E x} \\ \\   \, {b}^{e}_{\downarrow} \,e^{-i k^{e}_E x}\end{pmatrix}, \, \hspace{0.5cm} x <x_1,
\end{equation}
and
\begin{equation}\label{uNR}
\begin{pmatrix} u_{E\uparrow}(x) \\ \\ u_{E\downarrow}(x) \end{pmatrix} =   \begin{pmatrix}{b}^{e}_{\uparrow}\, e^{+i k^{e}_E x} \\ \\ {a}^{e}_{\downarrow} \, e^{-i k^{e}_E x}\end{pmatrix}, \, \hspace{0.5cm} x >x_2 ,
\end{equation}
respectively, where $k^e_E$ is given in Eq.(\ref{k^{e/h}_E}). For the wavefunction inside the barrier ($x_1 < x < x_2$), one has to distinguish two energy ranges.  
For $|\mu+E|<m_{\perp}$ the wavefunction consists of evanescent waves and reads
\begin{eqnarray}
\begin{pmatrix}
u_{E\uparrow}(x) \\ \\ u_{E\downarrow}(x) \end{pmatrix} &=& \frac{1}{\sqrt{2}}  \begin{pmatrix} e^{+\frac{i}{2} (\tilde{\theta}^{e}_E-\phi_{\perp})} \\ \\  e^{-\frac{i}{2} (\tilde{\theta}^{e}_E-\phi_{\perp})}  \end{pmatrix} \, e^{+i \tilde{k}^{e+}_E x} \,  {c}^{e}_{+} +    \nonumber \\
& & +\frac{1}{\sqrt{2}}  \begin{pmatrix} e^{-\frac{i}{2} (\tilde{\theta}^{e}_E+\phi_{\perp})} \\ \\  e^{+\frac{i}{2} (\tilde{\theta}^{e}_E+\phi_{\perp})}  \end{pmatrix} \, e^{+i \tilde{k}^{e-}_E x} \, {c}^{e}_{-}, \hspace{0.5cm}  \label{u-M-1}
\end{eqnarray}
where ${c}^{e}_{\pm}$ are complex amplitudes,  
\begin{equation}
\tilde{k}^{e\pm}_E=  \frac{-m_z \pm i \sqrt{m_{\perp}^2-(\mu+E)^2}}{\hbar v_F} \label{ktilde-p-def}
\end{equation}
are complex wavevectors, and $\tilde{\theta}^{e}_E$ is an angle such that
\begin{equation}
\begin{aligned}
    &\cos\tilde{\theta}^{e}_E = \frac{\mu+E}{m_{\perp}}, \\
    &\sin\tilde{\theta}^{e}_E = \frac{\sqrt{m_{\perp}^2-(\mu+E)^2}}{m_{\perp}}.
\end{aligned} \label{thetatilde-p-def}
\end{equation}
For $|\mu+E|>m_{\perp}$ the wavefunction exhibits an oscillatory behavior and is given by
\begin{eqnarray}
\lefteqn{\begin{pmatrix} u_{E\uparrow}(x) \\ \\ u_{E\downarrow}(x) \end{pmatrix}  =  } & &  \label{u-M-2}\\
&=& \sqrt{\frac{m_{\perp}}{ 2|\mu+E|}   }  \left(\begin{array}{l} e^{+\frac{1}{2} (s_E\bar{\theta}^{e}_E-i\phi_{\perp})} \\ \\  s_E e^{-\frac{1}{2} (s_E \bar{\theta}^{e}_E-i\phi_{\perp})}  \end{array} \right) \, e^{+i \bar{k}^{e+}_E x} \, {c}^{e}_{+} +     \nonumber \\
& & + \sqrt{\frac{m_{\perp}}{ 2|\mu+E|}   }\left(\begin{array}{l} e^{-\frac{1}{2} (s_E\bar{\theta}^{e}_E+i\phi_{\perp})} \\   \\ s_E e^{+\frac{1}{2} (s_E\bar{\theta}^{e}_E+i\phi_{\perp})}  \end{array} \right) \, e^{+i \bar{k}^{e-}_E x} \, {c}^{e}_{-}, \nonumber
\end{eqnarray}
where
\begin{equation}
\bar{k}^{e\pm}_E=  \frac{-m_z \pm   \sqrt{(\mu+E)^2-m_\perp^2}}{\hbar v_F}
\end{equation}
are real wavevectors, and $\bar{\theta}^{e}_E$ is determined by
\vskip -1em
\begin{equation}
\begin{aligned}
    &\cosh\bar{\theta}^{e}_E = \frac{|\mu+E|}{m_\perp}, \\
    &\sinh\bar{\theta}^{e}_E = \frac{\sqrt{(\mu+E)^2-m_\perp^2}}{m_\perp},
\end{aligned}
\end{equation}
with $s_E =\mbox{sgn}(\mu+E)$. 
Imposing the continuity at (say) the boundary $x_1$ of the wavefunctions in Eq.(\ref{u-NL}) and Eq.(\ref{u-M-1}) [or Eq.(\ref{u-M-2})], one obtains a set of two linear equations that lead to express the amplitudes ${c}^{e}_{\pm}$ characterizing the wavefunction inside the magnetic region as a function of the amplitudes $a^e_\uparrow$ and $b^e_\downarrow$. 
The above calculations straightforwardly enable  one to obtain the scattering matrix $S_e(E)$ of the barrier, which can be written in the form of Eq.(\ref{S0-e-gen}), where the transmission coefficient is 
\begin{widetext}
\begin{equation}
T_E = \left\{ \begin{array}{ll} 
\left(1+ \left(\frac{m_\perp}{\sqrt{m_\perp^2-(\mu+E)^2}} \sinh \left[ \frac{L_m}{\hbar v_F} \sqrt{m_\perp^2-(\mu+E)^2}\right]  \right)^2  \right)^{-1} & |\mu+E| < m_\perp, \\ & \\
\left(1+ \left(\frac{m_\perp}{\sqrt{(\mu+E)^2-m_\perp^2}} \sin  \left[ \frac{L_m}{\hbar v_F} \sqrt{(\mu+E)^2-m_\perp^2}\right]  \right)^2  \right)^{-1} & |\mu+E| >m_\perp, 
\end{array}\right. \label{TE-SGL}
\end{equation}
whereas the phases $\Gamma_m(E)$, $\Theta_m(E)$ and $\chi_m(E)$ are given by
\begin{equation}
\Gamma_m(E)= \left\{ \begin{array}{ll} 
\arctan \left[ \frac{\mu+E}{\sqrt{m_\perp^2-(\mu+E)^2}} \tanh\left( \frac{L_m}{\hbar v_F} \sqrt{m_\perp^2-(\mu+E)^2}\right)\right] \,- \frac{(\mu+E) L_m}{\hbar v_F} \hspace{1cm} & |\mu+E| < m_\perp, \\ & \\
\begin{array}{l}\arctan \left[ \frac{\mu+E}{\sqrt{(\mu+E)^2-m_\perp^2 }} \tan\left( \frac{L_m}{\hbar v_F} \sqrt{(\mu+E)^2-m_\perp^2}\right) \right]\, - \frac{(\mu+E) L_m}{\hbar v_F} + \hspace{1cm} \\
\hspace{1cm} + \pi\,\vartheta\left[-\cos\left( \frac{L_m}{\hbar v_F} \sqrt{(\mu+E)^2-m_\perp^2}\right)\right] 
\end{array} & |\mu+E| >m_\perp, 
\end{array}\right. \label{Gammam}
\end{equation}
and
\begin{eqnarray}
    \Theta_m(E)&=& 2 k_E^e x_0 +\phi_\perp,\\
    \chi_m(E) &=& \frac{m_z L_m}{\hbar v_F},
\end{eqnarray} 
respectively. Here, $x_0=(x_1+x_2)/2$ denotes the mid-point of the barrier,  $L_m=x_2-x_1$ is its spatial extension, and $\vartheta$ the Heaviside step function.

\end{widetext}

\bibliography{biblio} 
 
\end{document}